\documentclass[aps,pra,reprint,showkeys,superscriptaddress,amsmath,amssymb,final]{revtex4-2}
\usepackage[utf8]{inputenc}
\usepackage[T1]{fontenc}
\DeclareUnicodeCharacter{2212}{-}

\usepackage{graphicx}
\usepackage{xcolor}
\usepackage[caption=false]{subfig}
\newcommand{\phantomsubfloat}[1]{{
    \captionsetup[subfigure]{labelformat=empty}
    \subfloat[][]{#1}
}}
\usepackage{mathtools}
\usepackage{bm}
\usepackage[ISO]{diffcoeff}
\usepackage{braket}
\usepackage{siunitx}
\usepackage[colorlinks=true]{hyperref}

\newcommand{\keio}{School of Fundamental Science and Technology, Keio University, 3-14-1 Hiyoshi, Kohoku-ku, Yokohama, Kanagawa 223-8522 Japan}
\newcommand{\keioSCRN}{Center for Spintronics Research Network, Keio University, 3-14-1 Hiyoshi, Kohoku-ku, Yokohama, Kanagawa 223-8522 Japan}

\newcommand{\ketbra}[2]{\ket{#1}\,\bra{#2}}

\begin{document}
\preprint{APS/123-QED}

\title{Electron-spin double resonance of nitrogen-vacancy centers in diamond \\under strong driving field}
\date{\today}
\author{Takumi Mikawa}
    \email[Email: ]{mikawa\_jukusei@keio.jp}
    \affiliation{\keio}
    \affiliation{\keioSCRN}
\author{Ryusei Okaniwa}
    \affiliation{\keio}
    \affiliation{\keioSCRN}
\author{Yuichiro Matsuzaki}
    \email[Email: ]{matsuzaki.yuichiro@aist.go.jp}
    \affiliation{Research Center for Emerging Computing Technologies, National Institute of Advanced Industrial Science and Technology (AIST), 1-1-1 Umezono, Tsukuba, Ibaraki 305-8568, Japan}
    \affiliation{NEC-AIST Quantum Technology Cooperative Research Laboratory, National Institute of Advanced Industrial Science and Technology(AIST), 1-1-1 Umezono, Tsukuba, Ibaraki 305-8568, Japan}
\author{Norio Tokuda}
    \affiliation{Graduate School of Natural Science and Technology, Kanazawa University, Kakuma, Kanazawa, Ishikawa 920-1192, Japan}
    \affiliation{Nanomaterials Research Institute, Kanazawa University, Kakuma, Kanazawa, Ishikawa 920-1192, Japan}
\author{Junko Ishi-Hayase}
    \email[Email: ]{hayase@appi.keio.ac.jp}
    \affiliation{\keio}
    \affiliation{\keioSCRN}

\begin{abstract}
    The nitrogen-vacancy (NV) center in diamond has been the focus of research efforts because of its suitability for use in applications such as quantum sensing and quantum simulations.
    Recently, the electron-spin double resonance (ESDR) of NV centers has been exploited for detecting radio-frequency (RF) fields with continuous-wave optically detected magnetic resonance.
    However, the characteristic phenomenon of ESDR under a strong RF field remains to be fully elucidated.
    In this study, we theoretically and experimentally analyzed the ESDR spectra under strong RF fields by adopting the Floquet theory.
    Our analytical and numerical calculations could reproduce the ESDR spectra obtained by measuring the spin-dependent photoluminescence under the continuous application of microwaves and an RF field for a DC bias magnetic field perpendicular to the NV axis.
    We found that anticrossing structures that appear under a strong RF field are induced by the generation of RF-dressed states owing to the two-RF-photon resonances.
    Moreover, we found that $2n$-RF-photon resonances were allowed by an unintentional DC bias magnetic field parallel to the NV axis.
    These results should help in the realization of precise MHz-range AC magnetometry with a wide dynamic range beyond the rotating wave approximation regime as well as Floquet engineering in open quantum systems.
\end{abstract}

\maketitle

\section{Introduction}
    A nitrogen-vacancy (NV) center in diamond is a point defect composed of a substitutional nitrogen atom adjacent to a vacancy in the carbon lattice~\cite{Levine2019, Barry2020}.
    The electronic spin states of an NV center can be initialized by the illumination of a green laser and readout by measuring the spin-dependent photoluminescence.
    Moreover, the spin state can be manipulated by irradiating microwaves (MW) and exhibits a long coherence time even at room temperature.
    Owing to these properties, the NV center is a promising system for realizing quantum sensors with high sensitivity and spatial resolution~\cite{Ku2020, Huxter2022, Huxter2023} as well as feasible quantum simulators~\cite{Boyers2020, Zhang2021, Yang2022} under ambient conditions.

    Quantum sensing and simulation based on NV centers have been demonstrated using optically detected magnetic resonance (ODMR).
    ODMR has been performed using continuous-wave techniques (CW-ODMR)~\cite{Schloss2018, Tsukamoto2021, Chen2022} and pulsed techniques~\cite{Simon2017, Boss2017, Hart2021, Wang2022}.
    In the case of AC magnetic field sensing, several types of pulsed  (such as Hahn echo) techniques have been used.
    However, these techniques suffer from control errors and require careful calibration before measurements~\cite{Saijo2018, Yamaguchi2019, Tabuchi2023}.
    In contrast, the CW-ODMR technique is simple and employed widely because it uses continuous laser illumination and MW irradiation and does not require pulse control or careful calibration.
    However, the detectable frequency of CW-ODMR-based magnetometry is limited to values typically lower than the \si{\kilo\hertz} range.

    Recently, we proposed and successfully demonstrated MHz-range AC magnetometry using CW-ODMR~\cite{Saijo2018, Yamaguchi2019} by exploiting the phenomenon of electron-spin double resonance (ESDR) of NV centers~\cite{Saijo2018, Yamaguchi2019, Dmitriev2018, Dmitriev2019, Dmitriev2020, Dmitriev2022, Tabuchi2023}.
    During the ESDR measurements, we simultaneously and continuously irradiated a radio frequency (RF) field as the target and MW as the probe.
    If the RF field is coherently coupled to the electron spin states of the NV center, then RF-dressed states are generated, allowing one to probe the states by measuring the CW-ODMR spectrum by sweeping the MW frequencies under an RF field (ESDR spectrum).
    Thus, ESDR allows for the detection of MHz-range AC magnetic fields without pulse control and high-speed measurements.
    However, despite these advantages, the characteristic phenomenon of ESDR under a strong RF field remains to be completely elucidated~\cite{Saijo2018, Yamaguchi2019, Dmitriev2018, Dmitriev2019, Dmitriev2020, Dmitriev2022, Tabuchi2023}.
    Previous studies of ESDR focused on measuring weak RF fields.
    In the theoretical analysis performed in these studies, the rotating wave approximation (RWA), which is valid only for weak RF fields, played an important role~\cite{Saijo2018, Dmitriev2019, Yamaguchi2019, Tabuchi2023}.

    In the present study, we theoretically and experimentally analyzed the ESDR spectra of NV centers under strong RF fields.
    When driving a system with a strong external field, it is difficult to perform a theoretical analysis because the RWA is violated~\cite{Fuchs2009, London2014, Scheuer2014, Laucht2016}.
    To solve this problem, we adopted the Floquet theory~\cite{Shirley1965, Sambe1973}, which can be used to treat quantum systems driven by the time-periodic Hamiltonian~\cite{Eckardt2015, Oka2019, Ivanov2021}.
    The Floquet theory provides more precise solutions beyond the RWA regime, can account for muti-photon transitions, and yields further insights~\cite{Son2009, Hausinger2010, Deng2015, Han2019, Wang2021a, Nishimura2022}.
    Using the Floquet theory, we calculated the ESDR spectra under both weak and strong RF fields.
    Moreover, we performed ESDR experiments and found that the numerical and experimental results were in good agreement.
    Thus, our results should aid the realization of more precise MHz-range AC magnetometry with a wider dynamic range using CW-OMDR, fast and precise quantum control beyond the RWA regime~\cite{London2014, Scheuer2014, Wang2021a, Nishimura2022}, and  simulations of various quantum phenomena under strong driving fields~\cite{Martin2017, Boyers2020, Chen2020}.
    In particular, the NV center can be treated as a single-body system resistant to dissipation and is expected to serve as a platform for the Floquet engineering in open quantum systems~\cite{Ikeda2020, Ikeda2021, Mori2023}.

\section{Theory}
\subsection{Single RF-photon resonances}\label{sec:signle RF-photon}
    \begin{figure}[t]
        \centering
        \includegraphics[width=8.6cm]{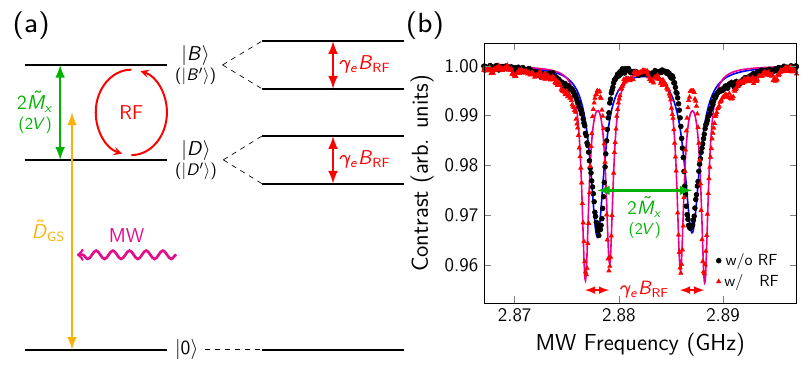}
        \phantomsubfloat{\label{subfigure 1a}}
        \phantomsubfloat{\label{subfigure 1b}}
        \label{figure 1}
        \caption{%
            (a) Energy diagram of electronic spin-triplet ground state of NV center under DC bias magnetic field perpendicular to NV axis under continuous application of MW and RF field.
            $\tilde{D}_{\mathrm{GS}}$ is zero-field splitting, and $\tilde{M}_x$ is effective strain under perpendicular magnetic field.
            (b) CW-ODMR spectra without and with the radio-frequency (RF) fields for the amplitude $B_\mathrm{RF} = \SI{69.8}{\micro\tesla}$ and frequency $\omega_\mathrm{RF}/(2\pi) = \SI{9.09}{\mega\hertz}$.
            In CW-ODMR spectrum under RF field (i.e., ESDR spectrum), four dips are observed owing to creation of RF-dressed states [see (a)].
            Solid lines indicate fitting curves calculated using multiple harmonic oscillator model.
        }
    \end{figure}
    In this section, we review the previous work about ESDR under a weak RF field in Ref.~\cite{Saijo2018, Yamaguchi2019}.
    We applied a DC bias magnetic field perpendicular to the NV axis~\cite{Shin2013}.
    It should be noted that the bias magnetic field contributes to suppressing inhomogeneous broadening owing to random magnetic fields.
    For $M_x \ll \gamma_e B_x, \gamma_e B_y \ll D_\mathrm{GS}$, the electronic spin-triplet (S = 1) ground state Hamiltonian of the NV center without an external oscillating field can be written as ($\hbar=1$) follows:
    \begin{equation}
        \hat{H}_0 \approx \tilde{D}_\mathrm{GS} \hat{S}_z^2 + \tilde{M}_x(\hat{S}_x^2 - \hat{S}_y^2)
        \label{eq:Hamiltonian under perpendicular bias magnetic field}
    \end{equation}
    where $\bm{\hat{S}}=(\hat{S}_x,\hat{S}_y,\hat{S}_z)$ is the dimensionless spin-1 operators for the electronic spin.
    Here, we consider the direction of the effective strain to be the $x$ direction.
    Furhermore, $\tilde{D}_{\mathrm{GS}}$ is the zero-field splitting, and $\tilde{M}_x$ is the effective strain under the perpendicular magnetic field~\cite{Yamaguchi2019}.
    Under the perpendicular DC bias field, the eigenstates can be approximated as $\ket{B}\coloneqq(\ket{m_s=+1}+\ket{m_s=-1})/\sqrt{2}$ and  $\ket{D}\coloneqq(\ket{m_s=+1}-\ket{m_s=-1})/\sqrt{2}$ and $\ket{0}$~\cite{Matsuzaki2016, Saijo2018, Yamaguchi2019}.
    The energy diagram is shown in Fig.~\ref{subfigure 1a}.

    For the ESDR, we assumed simultaneous irradiation with MW and an RF field.
    In this case, the Hamiltonian is given by
    \begin{equation}
        \begin{aligned}
            \hat{H}(t) &\approx \tilde{D}_\mathrm{GS} \hat{S}_z^2 + \tilde{M}_x(\hat{S}_x^2 - \hat{S}_y^2) \\
                       &\qquad + (2\lambda^b\hat{S}_x + 2\lambda^d\hat{S}_y)\cos{\omega_\mathrm{MW}t} \\
                       &\qquad\qquad + 2\Omega_\mathrm{RF}\hat{S}_z\cos{\omega_\mathrm{RF}t}
        \end{aligned}
    \end{equation}
    where $\lambda^b$ ($\lambda^d$) is the amplitude of the MW in the $x$ ($y$) direction, $\Omega_\mathrm{RF}$ is the amplitude of the RF field in the $z$ direction, and $\omega_\mathrm{MW}$ ($\omega_\mathrm{RF}$) is the MW (RF) frequency.
    Here, we ignore the longitudinal components of the MW and the transverse components of the RF field because they oscillate with a high frequency without any resonance.
    In addition, we move to a rotating frame defined by $\hat{R}_0(t) = e^{-i\omega_\mathrm{MW}\hat{S}_z^2}$.
    By using the RWA, we can rewrite this Hamiltonian as
    \begin{equation}
        \begin{aligned}
            \hat{H}(t) &\approx \Delta_\mathrm{MW} \hat{S}_z^2 + \tilde{M}_x(\hat{S}_x^2 - \hat{S}_y^2) \\
                       &\qquad + \lambda^b\hat{S}_x + \lambda^d\hat{S}_y + 2\Omega_\mathrm{RF}\hat{S}_z \cos{\omega_\mathrm{RF}t}
        \end{aligned}
        \label{eq:the Hamiltonian of the ESDR in the first rotating frame with the RWA}
    \end{equation}
    where $\Delta_\mathrm{MW} \coloneqq \tilde{D}_\mathrm{GS} - \omega_\mathrm{MW}$ is detuning.
    To derive Eq.~\eqref{eq:the Hamiltonian of the ESDR in the first rotating frame with the RWA}, we drop the high-oscillation term, that is, $e^{\pm 2i \omega_\mathrm{MW} t}$.
    Because $2\omega_\mathrm{MW}$ is significantly larger than any of the parameters in Eq.~\eqref{eq:the Hamiltonian of the ESDR in the first rotating frame with the RWA} except the zero-field splitting, $\tilde{D}_\mathrm{GS}$, we assumed that this approximation would be valid throughout this study.

    The Hamiltonian in Eq.~\eqref{eq:the Hamiltonian of the ESDR in the first rotating frame with the RWA} is periodic in time.
    Thus, we can adopt the Floquet theory to describe the dynamics beyond the RWA regime.
    By using the Floquet theory, we obtain a time-independent Hamiltonian instead of a time-periodic one (See Appendix~\ref{app:Floquet theory} for details).
    To simplify the notation, we use the extended Hilbert space or the Sambe space, $\mathfrak{F}$~\cite{Sambe1973}.
    Then, we can express the time-independent Hamiltonian using the basis of $\ket{\alpha, n} = \ket{\alpha} \otimes \ket{n}$, where $\alpha$ is the state of the NV center, and $n$ is the Fourier index.
    In this case, the Fourier index, $n$ can be interpreted as the number of absorbed RF photons (if $n$ is negative, $|n|=-n$ can be interpreted as the number of emitted RF photons)~\cite{Eckardt2015}.
    Thus, we can treat $\ket{n}$ as the Fock state.
    In the Sambe space, the quantum dynamics can be described using the Floquet Hamiltonian $\hat{H}_F$ as follows
    \begin{align}
            \hat{H}_F &\coloneqq \sum_m \hat{\mathcal{F}}_m \otimes \hat{H}^{(m)} + \hat{\mathcal{N}} \otimes \omega_\mathrm{RF}\hat{1}
    \end{align}
    where $\hat{\mathcal{F}}_n$ are the Floquet ladder operators, $\hat{\mathcal{N}}$ is  the Floquet number operator, and $\hat{H}^{(m)}$ is the Fourier coefficient.
    These are defined as follows~\cite{Ivanov2021}:
    \begin{align}
        \hat{\mathcal{F}}_n\ket{m} = \ket{n+m}, \\
        \hat{\mathcal{N}}\ket{n} = n\ket{n}, \\
        \hat{H}(t) = \sum_m \hat{H}^{(m)}e^{i m \omega_\mathrm{RF} t}.
    \end{align}
    For the ESDR, the Floquet Hamiltonian can be represented by the matrix using the basis of $\{\ket{B}, \ket{0}, \ket{D}\}$ as follows:
    \begin{widetext}
        \renewcommand\arraystretch{1.2}
        \begin{align}
            &{} \hat{H}_F = \notag \\
            &\left[
            \begin{array}{c|ccc|ccc|ccc|c}
                \ddots & \vdots & \vdots & \vdots & \vdots & \vdots & \vdots & & & & \\
                \hline
                \cdots & \Delta_\mathrm{MW}^b + \omega_\mathrm{RF} & \lambda^b & 0 & 0 & 0 & \Omega_\mathrm{RF} & & & & \\
                \cdots & \lambda^b & \omega_\mathrm{RF} & i\lambda^d & 0 & 0 & 0 & & & & \\
                \cdots & 0 & -i\lambda^d & \Delta_\mathrm{MW}^d + \omega_\mathrm{RF} & \Omega_\mathrm{RF} & 0 & 0 & & & & \\
                \hline
                \cdots & 0 & 0 & \Omega_\mathrm{RF} & \Delta_\mathrm{MW}^b & \lambda^b & 0 & 0 & 0 & \Omega_\mathrm{RF} & \cdots \\
                \cdots & 0 & 0 & 0 &\lambda^b & 0 & i\lambda^d & 0 & 0 & 0 & \cdots \\
                \cdots &\Omega_\mathrm{RF} & 0 & 0 & 0 & -i\lambda^d & \Delta_\mathrm{MW}^d & \Omega_\mathrm{RF} & 0 & 0 & \cdots \\
                \hline
                & & & & 0 & 0 & \Omega_\mathrm{RF} & \Delta_\mathrm{MW}^b - \omega_\mathrm{RF} & \lambda^b & 0 & \cdots \\
                & & & & 0 & 0 & 0 &\lambda^b & -\omega_\mathrm{RF} & i\lambda^d & \cdots \\
                & & & & \Omega_\mathrm{RF} & 0 & 0 & 0 & -i\lambda^d & \Delta_\mathrm{MW}^d - \omega_\mathrm{RF} & \cdots \\
                \hline
                & & &  & \vdots & \vdots & \vdots & \vdots & \vdots & \vdots & \ddots
            \end{array}
            \right]
            \begin{array}{cc}
                 & \vdots \\
                \leftarrow & \ket{B,+1} \\
                \leftarrow & \ket{0,+1} \\
                \leftarrow & \ket{D,+1} \\
                \leftarrow & \ket{B,0} \\
                \leftarrow & \ket{0,0} \\
                \leftarrow & \ket{D,0} \\
                \leftarrow & \ket{B,-1} \\
                \leftarrow & \ket{0,-1} \\
                \leftarrow & \ket{D,-1} \\
                 & \vdots
            \end{array}
            \label{eq:Floquet Hamiltonian for ESDR}
        \end{align}
    \end{widetext}
    where $\Delta_\mathrm{MW}^b \coloneqq \Delta_\mathrm{MW} + \tilde{M}_x,\Delta_\mathrm{MW}^d \coloneqq \Delta_\mathrm{MW} - \tilde{M}_x$.

    The time-averaged transition probability from  $\ket{\alpha}$ to $\ket{\beta}$ is calculated as follows:
    \begin{equation}
        \braket{P_{\alpha\to\beta}} = \sum_{n,m}\sum_k |\braket{\beta,n|q_k,m}|^2|\braket{q_k,m|\alpha,0}|^2
        \label{eq:averaged transition probability in Floquet theory}
    \end{equation}
    where $\ket{q_k,m}$ are the eigenstates of the Floquet Hamiltonian~\cite{Shirley1965}.
    To obtain  $\ket{q_k,m}$, we must solve the infinite-dimensional eigenvalue equation (See Eqs.~\eqref{eq:eigenvalue equation of Floquet Hamiltonian} and \eqref{eq:Floquet-Schrodinger equation} for the details).
    Specifically, by truncating the Floquet Hamiltonian to a finite size, we can calculate the transition probability through a numerical simulation.

    When the ESDR condition $|\omega_\mathrm{RF} - 2\tilde{M}_x| \ll \Omega_\mathrm{RF}$ is satisfied, $\ket{B,n}$ and $\ket{D,n+1}$ can be coupled to the RF field, and a single-RF-photon resonance occurs.
    Focusing on the single-RF-photon resonance (e.g., that between $\ket{B,0}$ and $\ket{D,+1}$) we can reduce the infinite-dimensional  Floquet Hamiltonian in Eq.~\eqref{eq:Floquet Hamiltonian for ESDR} a $3 \times 3$ effective Hamiltonian:
    \begin{equation}
        \hat{H}_\mathrm{eff} \approx
            \begin{bmatrix}
                \Delta_\mathrm{MW}^b & \lambda^b & \Omega_\mathrm{RF} \\
                \lambda^b & 0 & 0 \\
                \Omega_\mathrm{RF} & 0 & \Delta_\mathrm{MW}^d + \omega_\mathrm{RF}
            \end{bmatrix}.
    \end{equation}
    This matrix can also be derived by the RWA in the rotating frame defined by $e^{-i\omega_\mathrm{RF}t\ketbra{D}{D}}$, as in Ref.~\cite{Yamaguchi2019}.
    Using Fermi's golden rule or the harmonic oscillator model~\cite{Matsuzaki2016, Yamaguchi2019, Tabuchi2023} under the weak driving condition $\Omega_\mathrm{RF} \ll \omega_\mathrm{RF}$, we can obtain the resonant MW frequencies~\cite{Saijo2018, Yamaguchi2019}, as follows:
    \begin{equation}
        \omega_\mathrm{MW} = \tilde{D}_\mathrm{GS} \pm \frac{1}{2}\omega_\mathrm{RF} \pm \sqrt{\left(\tilde{M}_x - \frac{1}{2}\omega_\mathrm{RF}\right)^2 + \Omega_\mathrm{RF}^2}.
        \label{eq:resonant MW frequencies of single-RF-photon resonance}
    \end{equation}
    Equation~\eqref{eq:resonant MW frequencies of single-RF-photon resonance} indicates that the anticrossings occur at the RF frequency $\omega_\mathrm{RF} \approx 2\tilde{M}_x$.
    These anticrossing gaps are called the Aulter–Towns splittings~\cite{Morishita2019, Dmitriev2019, Yamaguchi2019} and are proportional to the RF amplitude $\Omega_\mathrm{RF}$.
    Thus, upon the irradiation of a resonant RF field, we observed  four dips in the CW-ODMR spectrum (i.e., ESDR spectrum) with the Aulter–Towns splittings $2\Omega_\mathrm{RF} = \gamma_e B_\mathrm{RF}$ (See Fig.~\ref{subfigure 1b}), which allowed for MHz-range AC magnetic field sensing~\cite{Saijo2018, Yamaguchi2019}.

\subsection{Multi-RF-photon resonances}\label{sec:multi RF-photon resonance}
    Next, we consider  ESDR under a strong RF field $\Omega_\mathrm{RF} \gtrsim \omega_\mathrm{RF}$, which results in multi-RF-photon resonances.
    The simple model for using the RWA~\cite{Saijo2018, Yamaguchi2019} does not explain these phenomena, while the Floquet theory can reproduce the experimental results in this regime, as described later.
    In the case of a single-RF-photon resonance, the non-zero off-diagonal terms with $\Omega_\mathrm{RF} \ne 0$ in the Floquet Hamiltonian induce the transition between $\ket{B,n}$ and $\ket{D,n+1}$.
    Here, the RF photon number, $n$, changes with $\Delta m = +1$.
    On the other hand, other transitions also occur between the states with different RF photon numbers in the case of  multi-RF-photon resonances.
    The key point is that even when the off-diagonal terms between the states in the Floquet Hamiltonian are zero, the transitions between the specific states can occur via indirect transitions that use the other states as the intermediate states.

    Let us consider an example of one such indirect transition owing to strong RF driving.
    In Eq.~\eqref{eq:Floquet Hamiltonian for ESDR}, there is an indirect transition between $\ket{B, -1}$ and $\ket{B,+1}$.
    Let us focus on this transition.
    While there are no direct transitions between $\ket{B, -1}$ and $\ket{B,+1}$, we can induce a transition from $\ket{B,-1}$ to $\ket{D,0}$ and subsequently induce a transition from $\ket{D,0}$ to $\ket{B,+1}$.
    In this case, the quantum number, $n$, changes by $+2$, which corresponds to the two-RF-photon transitions.
    It should be noted that, during the transition described above, the spin state does not change, thus the so-called anticrossing is not observed during the spectroscopy.
    In addition, more RF-photon transitions can occur if we consider higher-order transitions, including three-RF-photon transitions through a sequence of transitions, such as $\ket{B,-2} \to \ket{D,-1} \to \ket{B,0} \to \ket{D,+1}$.
    In this case, an anticrossing structure should be observed since the spin state is changed.
    By performing similar calculations, we can show that $2n$-RF-photon transitions from $|B,-1\rangle $ to $|B,2n-1\rangle $ cannot induce the anticrossing structures while $(2n-1)$-RF-photon transitions from $|B,-2\rangle $ to $|D,2n-1\rangle$ can.

    Importantly, if we consider the effect of the DC bias magnetic field parallel to the NV axis, $\omega_L \coloneqq \gamma_e B^{(\mathrm{bias})}_z$, additional anticrossing structures occur owing to the multi-RF-photon resonances.
    Such magnetic fields may originate because of the misalignment of the perpendicular bias magnetic field or the Earth's magnetic field ($\approx \SI{0.05}{\milli\tesla}$).
    The Hamiltonian under the bias magnetic field in the absence of drivings is given as
    \begin{equation}
        \hat{H}_0' = \tilde{D}_\mathrm{GS}\hat{S}_z^2 + \tilde{M}_x(\hat{S}_x^2 - \hat{S}_y^2) + \omega_L\hat{S}_z.
    \end{equation}
    The lowest-energy eigenstate of this Hamiltonian is $|0\rangle$ with an eigenenergy of $0$.
    The other eigenstates of this Hamiltonian are as follows:
    \begin{equation}
        \begin{gathered}
            \ket{B'} \coloneqq \frac{(\tilde{M}_x + V)\ket{B} + \omega_L\ket{D}}{\sqrt{(\tilde{M}_x + V)^2 + \omega_L^2}}, \\
            \ket{D'} \coloneqq \frac{-\omega_L\ket{B} + (\tilde{M}_x + V)\ket{D}}{\sqrt{(\tilde{M}_x + V)^2 + \omega_L^2}}
        \end{gathered}
    \end{equation}
    with eigenvalues of $\tilde{D}_\mathrm{GS} + V$ and $\tilde{D}_\mathrm{GS} - V$, respectively where $V \coloneqq \sqrt{\tilde{M}_x^2 + \omega_L^2}$.
    By using this basis, we can rewrite the Hamiltonian of the NV center, including the MW and the RF field in the rotating frame, which is defined by $\hat{R}_0(t)$ with the RWA as
    \begin{widetext}
        \begin{equation}
            \hat{H}'(t) =
                \begin{bmatrix}
                        \Delta_\mathrm{MW} + V + 2\dfrac{\omega_L\Omega_\mathrm{RF}}{V}\cos{\omega_\mathrm{RF}t} &  \lambda^{b'} & 2\dfrac{\tilde{M}_x\Omega_\mathrm{RF}}{V}\cos{\omega_\mathrm{RF}t} \\
                         (\lambda^{b'})^* & 0 & i\lambda^{d'} \\
                         2\dfrac{\tilde{M}_x\Omega_\mathrm{RF}}{V}\cos{\omega_\mathrm{RF}t} & -i(\lambda^{d'})^* & \Delta_\mathrm{MW} - V - 2\dfrac{\omega_L\Omega_\mathrm{RF}}{V}\cos{\omega_\mathrm{RF}t}
                \end{bmatrix}
            \label{eq:Hamiltonian for ESDR in new basis}
        \end{equation}
        where $\lambda^{b'}$ ($\lambda^{d'}$) is the effective MW amplitude corresponding to the transition between $\ket{0}$ and $\ket{B'}$ ($\ket{D'}$).
        From the Fourier expansion, the Floquet Hamiltonian for the ESDR can be represented as follows:
        \renewcommand\arraystretch{1.5}
        \begin{align}
            &\hat{H}_F' = \notag \\
            &\left[
            \begin{array}{c|ccc|ccc|ccc|c}
                \ddots & \vdots & \vdots & \vdots & \vdots & \vdots & \vdots & & & & \\
                \hline
                \cdots & \Delta_1^{b'} & \lambda^{b'} & 0 & \omega_L \Omega_\mathrm{RF}/V & 0 & \tilde{M}_x\Omega_\mathrm{RF}/V & & & & \\
                \cdots & (\lambda^{b'})^* & \omega_\mathrm{RF} & i\lambda^{d'} & 0 & 0 & 0 & & & & \\
                \cdots & 0 & (i\lambda^{d'})^* & \Delta_1^{d'} & \tilde{M}_x\Omega_\mathrm{RF}/V & 0 & -\omega_L \Omega_\mathrm{RF}/V & & & & \\
                \hline
                \cdots & \omega_L \Omega_\mathrm{RF}/V & 0 & \tilde{M}_x\Omega_\mathrm{RF}/V & \Delta_0^{b'} & \lambda^{b'} & 0 & \omega_L \Omega_\mathrm{RF}/V & 0 & \tilde{M}_x\Omega_\mathrm{RF}/V & \cdots \\
                \cdots & 0 & 0 & 0 & (\lambda^{b'})^* & 0 & i\lambda^{d'} & 0 & 0 & 0 & \cdots \\
                \cdots & \tilde{M}_x\Omega_\mathrm{RF}/V & 0 & -\omega_L \Omega_\mathrm{RF}/V & 0 & (i\lambda^{d'})^* & \Delta_0^{d'} & \tilde{M}_x\Omega_\mathrm{RF}/V & 0 & -\omega_L \Omega_\mathrm{RF}/V & \cdots \\
                \hline
                & & & & \omega_L \Omega_\mathrm{RF}/V & 0 & \tilde{M}_x\Omega_\mathrm{RF}/V & \Delta_{-1}^{b'} & \lambda^{b'} & 0 & \cdots \\
                & & & & 0 & 0 & 0 & (\lambda^{b'})^* & -\omega_\mathrm{RF} & i\lambda^{d'} & \cdots \\
                & & & & \tilde{M}_x\Omega_\mathrm{RF}/V & 0 & -\omega_L \Omega_\mathrm{RF}/V & 0 & (i\lambda^{d'})^* & \Delta_{-1}^{d'} & \cdots \\
                \hline
                & & &  & \vdots & \vdots & \vdots & \vdots & \vdots & \vdots & \ddots
            \end{array}
            \right]
            \label{eq:Floquet Hamiltonian for ESDR with bias magnetic field}
        \end{align}
    \end{widetext}
    where $\Delta_n^{b'} \coloneqq \Delta_\mathrm{MW} + V +n\omega_\mathrm{RF}$ and $\Delta_n^{d'} \coloneqq \Delta_\mathrm{MW} - V + n\omega_\mathrm{RF}$.

    Then, the new off-diagonal terms, $\omega_L \Omega_\mathrm{RF}/V$ allow other subsequent transitions, such as $\ket{B,-1} \to \ket{B,0} \to \ket{D,+1}$.
    These subsequent transitions indicate that the anticrossing structures  owing to the two-RF-photon resonances are created approximately at half the RF frequency at which the single-RF-photon resonances occur.

\subsection{Analytical Solutions}\label{sec:Analytical Solutions}
    To obtain an approximate analytical solution for the resonant frequency of these two-RF-photon resonances, we use the Jacobi-Anger expansion~\cite{Childress2010, Meinel2021}.
    This method allows for the conversion of multistep single-photon transitions into direct multi-photon transitions.
    Before calculating the Floquet Hamiltonian of the ESDR, we move to a rotating frame with a siary operator defined as
    \begin{equation}
        \hat{U} = \exp\left[i \frac{2\omega_L \Omega_\mathrm{RF}}{\omega_\mathrm{RF} V}\sin{\omega_\mathrm{RF}t} (\ketbra{B'}{B'} - \ketbra{D'}{D'})\right].
    \end{equation}
    Then, the Hamiltonian in Eq.\eqref{eq:Hamiltonian for ESDR in new basis} becomes
    \begin{widetext}
        \begin{equation}
            \hat{H}_I(t) =
            \begin{bmatrix}
                \Delta_\mathrm{MW}^{b'} & \sum_k \lambda^{b'}_k e^{ik\omega_\mathrm{RF}t} & \sum_k \Omega_k e^{ik\omega_\mathrm{RF}t} \\
                \sum_k (\lambda^{b'}_k)^* e^{-ik\omega_\mathrm{RF}t} & 0 & i \sum_k \lambda^{d'}_k e^{ik\omega_\mathrm{RF}t} \\
                \sum_k \Omega_k e^{-ik\omega_\mathrm{RF}t} & -i \sum_k (\lambda^{d'}_k)^* e^{-ik\omega_\mathrm{RF}t} & \Delta_\mathrm{MW}^{d'}
            \end{bmatrix}
        \end{equation}
        where $\lambda^{b'}_k \coloneqq \lambda^{b'}J_k\left(\dfrac{2\omega_L\Omega_\mathrm{RF}}{\omega_\mathrm{RF}V}\right),\lambda^{d'}_k \coloneqq \lambda^{d'}J_k\left(\dfrac{2\omega_L\Omega_\mathrm{RF}}{\omega_\mathrm{RF}V}\right), \Omega_k \coloneqq \dfrac{k\omega_\mathrm{RF}\tilde{M}_x}{2\omega_L}J_k\left(\dfrac{4\omega_L\Omega_\mathrm{RF}}{\omega_\mathrm{RF}V}\right)$.
    \end{widetext}
    We used
    \begin{equation}
        \begin{gathered}
            e^{i\frac{x}{\omega}\sin{\omega t}} = \sum_{k=-\infty}^\infty J_k(x) e^{i k \omega t}, \\
            J_{k-1}(x) + J_{k+1}(x) = \frac{2k}{x}J_k(x).
        \end{gathered}
    \end{equation}
    where $J_k(x)$ is the $k$-th order Bessel function of the first kind.
    For the RF-dressed states generated by the coupling between $\ket{B,m}$ and $\ket{D,n}$, we move to the rotating frame defined by $e^{-i\omega_\mathrm{RF}t(m\ketbra{B'}{B'} + n\ketbra{D'}{D'})}$ and we use the RWA, wherein we ignore all the oscillating terms~\cite{Son2009, Hausinger2010, Han2020}.
    Then, we obtain the time-independent Hamiltonian as follows:
    \begin{equation}
        \hat{H}_\mathrm{RWA} \approx
            \begin{bmatrix}
                \Delta_m^{b'} & \lambda^{b'}_m & \Omega_{m-n} \\
                (\lambda^{b'}_m)^* & 0 & i \lambda^{d'}_{-n} \\
                \Omega_{m-n} & (i\lambda^{d'}_{-n})^* & \Delta_n^{d'}
            \end{bmatrix}.
        \label{eq:Hamiltonian of general RF-dressed states with RWA}
    \end{equation}
    Since the MW driving is weak, we assume that $\lambda_k^{b'}\simeq 0$ and $\lambda_k^{d'}\simeq 0$.
    In this case, we can analytically diagonalize the Hamiltonian in Eq.~\eqref{eq:Hamiltonian of general RF-dressed states with RWA}.
    Based on the energy difference between the ground and  excited state, we obtain the resonant frequency as follows:
    \begin{equation}
        \begin{aligned}
            \omega_\mathrm{MW} &\approx \tilde{D}_\mathrm{GS} + \frac{m+n}{2}\omega_\mathrm{RF} \\
                             &\quad \pm \sqrt{\left(V + \frac{m-n}{2}\omega_\mathrm{RF}\right)^2 + \Omega_{m-n}^2}.
        \end{aligned}
        \label{eq:resonant MW frequencies of multi-RF-photon resonance}
    \end{equation}
    When the resonant condition $\Delta^{b'}_m \approx \Delta^{d'}_n$ is satisfied, the $(m-n)$-RF-photon resonance occurs and generates the RF-dressed states, which exhibit an energy split of $2\Omega_{m-n}$.
    The RWA is valid when $\lambda^{b'}_k$ and $\lambda^{d'}_k$ are much smaller than $\omega_\mathrm{RF}$.
    In addition, as we increase $\Omega_\mathrm{RF}$, the Bessel functions, $J_k\left(\frac{2\omega_L\Omega_\mathrm{RF}}{\omega_\mathrm{RF}V}\right)$ and $J_k\left(\frac{4\omega_L\Omega_\mathrm{RF}}{\omega_\mathrm{RF}V}\right)$, become smaller, and the RWA becomes more accurate~\cite{Han2020}.
    However, since it is experimentally difficult to realize such an ultra-strong RF driving regime, confirming this was out of the scope of this study.

    RWA is valid when the off-diagonal component $\Omega_{m-n}$ is much smaller than the oscillating frequency, $\omega_\mathrm{RF}$.
    Therefore, as we increase the amplitude of the RF driving, the approximation breaks down.
    As a result, we cannot explain some of the resonances by using this analytical solution.
    On the other hand, the numerical results with the Floquet theory are still valid even for strong RF driving.

    To overcome the limitations of the RWA, we use the van Vleck (vV) transformation~\cite{Leskes2010, Ivanov2021, Wang2021b}.
    When calculating the effective Hamiltonian Eq.~\eqref{eq:Hamiltonian of general RF-dressed states with RWA} up to the second-order correction  using the vV transformation, we obtain
    \begin{equation}
        \begin{aligned}
            \hat{H}_\mathrm{vV}
                &\approx \hat{H}_\mathrm{vV}^{(1)} + \hat{H}_\mathrm{vV}^{(2)} \\
                &=
                    \begin{bmatrix}
                        \Delta_m^{b'} + \delta^{b'}_{m,n} & \lambda^{b'}_m & \Omega_{m-n} \\
                        (\lambda^{b'}_m)^* & \delta^{0}_{m,n} & i \lambda^{d'}_{-n} \\
                        \Omega_{m-n} & -i (\lambda^{d'}_{-n})^* & \Delta_n^{d'} + \delta^{d'}_{m,n}
                    \end{bmatrix}
        \end{aligned}
    \end{equation}
    where
    \begin{gather}
        \delta^{b'}_{m,n} \coloneqq \sum_{k\ne0} \frac{\Omega_{k-m+n}^2 - \Omega_{k+m-n}^2 + |\lambda^{b'}_{k-m}|^2 - |\lambda^{b'}_{k+m}|^2}{2k\omega_\mathrm{RF}} \\
        \delta^0_{m,n} \coloneqq \sum_{k\ne0} \frac{|\lambda^{b'}_{k+m}|^2 - |\lambda^{b'}_{k-m}|^2 - |\lambda^{d'}_{k+n}|^2 + |\lambda^{d'}_{k-n}|^2}{2k\omega_\mathrm{RF}} \\
        \delta^{d'}_{m,n} \coloneqq \sum_{k\ne0} \frac{\Omega_{k+m-n}^2 - \Omega_{k-m+n}^2 + |\lambda^{d'}_{k-n}|^2 - |\lambda^{d'}_{k+n}|^2}{2k\omega_\mathrm{RF}}.
    \end{gather}
    By calculating the energy difference between the ground and excited states, we obtain the resonant MW frequencies as
    \begin{equation}
        \begin{aligned}
            \omega_\mathrm{MW} &\approx \tilde{D}_\mathrm{GS} - \delta^0_{m,n} + \frac{m+n}{2}\omega_\mathrm{RF} \\
                               &\pm \sqrt{\left(V + \frac{\delta^{b'}_{m,n} - \delta^{d'}_{m,n}}{2} + \frac{m-n}{2}\omega_\mathrm{RF}\right)^2 + \Omega_{m-n}^2}.
        \end{aligned}
        \label{eq:resonant MW frequencies of multi-RF-photon resonance by van Vleck transformation}
    \end{equation}
    By including the higher-order corrections described in Eq.~\eqref{eq:Floquet Hamiltonian for ESDR with bias magnetic field}, we  can obtain a more accurate analytical solution.
    This is left for a future work study.

\section{Result}
\subsection{Setup}
    To verify this theory, we performed experiments using a home-built confocal laser microscope setup with NV ensembles, as in Ref.~\cite{Tabuchi2023}.
    The diamond sample used was a N-doped CVD-grown NV layer with a thickness of \SI{4.9}{\micro\meter} on a $(111)$-oriented diamond substrate.
    The NV concentration was estimated to be $\sim\SI{e16}{\per\centi\meter\cubed}$.
    The NV orientation was perferentially aligned along the $[111]$ direction of the diamond lattice~\cite{Michl2014, Lesik2014, Fukui2014, Ishiwata2017}.
    The NV ensembles were excited using a \SI{532}{\nano\meter} green laser with an average power of \SI{0.080}{\milli\watt}.
    The spin-dependent photoluminescence was measured using an avalanche photodiode under the continuous application of MW and RF fields to obtain the CW-ODMR spectra under  ESDR conditions (i.e., the ESDR spectra).
    The MW was irradiated at a power of \SI{-30}{\decibel} from an MW antenna placed on the opposite side of the NV layer~\cite{Sasaki2016}.
    The RF field was irradiated by placing a copper wire on the NV layer.
    During all the ESDR experiments performed in this study, we swept the MW frequency for different RF frequencies and amplitudes under a DC bias magnetic field perpendicular to the NV axis.

\subsection{Preliminary experiment without RF field}
    As a preliminary experiment, we measured the CW-ODMR spectrum without an RF field.
    As shown in Fig.~\ref{subfigure 1b}, we observed two dips corresponding to the two resonances.
    One of them corresponds to a transition from $\ket{0}$ to $\ket{B'}$, while the other corresponds to a transition from $\ket{0}$ to $\ket{D'}$.
    This ODMR spectrum could be fitted using a harmonic oscillator model~\cite{Matsuzaki2016, Yamaguchi2019, Tabuchi2023}.
    From the fitting, we obtained $\tilde{D}_\mathrm{GS}/(2\pi)=\SI{2.8825}{\giga\hertz}$ (zero-field splitting), $2V =\SI{9.09}{\mega\hertz}$ (the resonant frequency between $\ket{B'}$ and $\ket{D'}$), and $\lambda^b/(2\pi)=\lambda^d/(2\pi)=\SI{0.12}{\mega\hertz}$ (the MW amplitudes).

\subsection{Weak RF regime}
    \begin{figure}
        \centering
        \includegraphics[width=8.6cm]{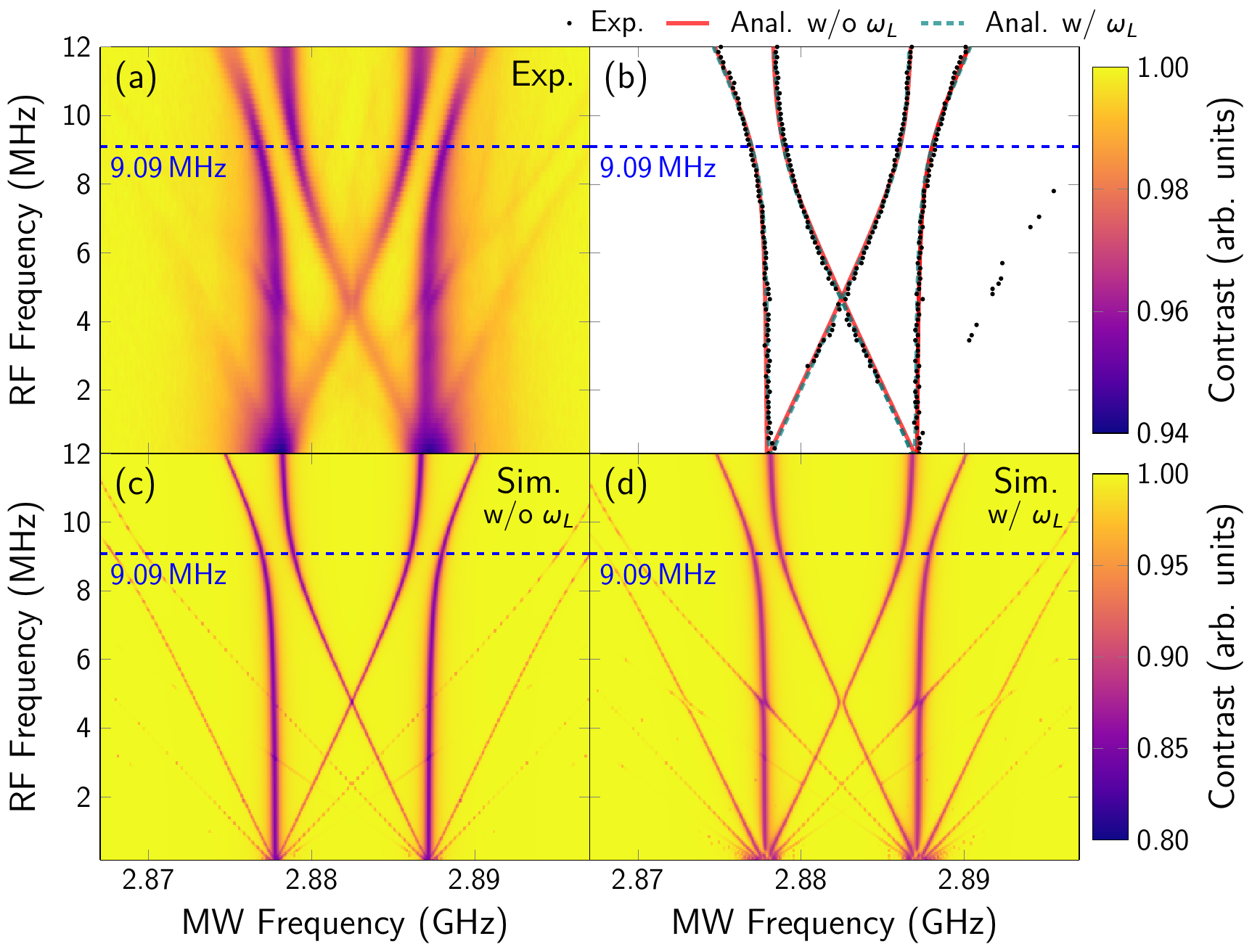}
        \phantomsubfloat{\label{subfigure 2a}}
        \phantomsubfloat{\label{subfigure 2b}}
        \phantomsubfloat{\label{subfigure 2c}}
        \phantomsubfloat{\label{subfigure 2d}}
        \caption{%
            (a) ESDR spectra under DC bias magnetic field perpendicular to NV axis as measured by sweeping MW and RF frequencies for a weak RF field with $B_\mathrm{RF} = \SI{69.8}{\micro\tesla}$.
            (b) Comparison of the resonant frequencies obtained experimentally and those obtained from analytical solutions under a weak RF field.
            Experimental resonant frequencies were extracted from (a), while analytical solutions without and with  parallel DC bias magnetic field $\omega_L$ were obtained using Eqs.~\eqref{eq:resonant MW frequencies of single-RF-photon resonance} and \eqref{eq:resonant MW frequencies of multi-RF-photon resonance}.
            (c) and (d) Results of numerical simulations performed using Floquet Hamiltonian (c) without and (d) with  parallel DC bias magnetic field.
            Parameters used for simulation in (c) [(d)] are $\tilde{D}_\mathrm{GS}/(2\pi) = \SI{2.8825}{\giga\hertz}$, $\tilde{M}_x/(2\pi) = \SI{9.09}{\mega\hertz}/2$ [$\tilde{M}_x/(2\pi) =  \SI{4.40}{\mega\hertz}$, $\omega_L/(2\pi) = \SI{0.50}{\mega\hertz}$] and $\lambda^{b'}/(2\pi) = \lambda^{d'}/(2\pi) =  \SI{0.12}{\mega\hertz}$.
            Dashed lines in (a) and (b) indicate resonant frequency between $\ket{B}$ ($\ket{B'}$) and $\ket{D}$ ($\ket{D'}$), anticrossing structures emerge near dashed lines.
            There is good agreement between experimental and analytical results, and parallel bias magnetic field had no effect under the weak RF field.
        }
        \label{figure 2}
    \end{figure}
    Next, we performed the ESDR experiment under a weak RF field with $\Omega_\mathrm{RF}/(2\pi) \approx \SI{1.96}{\mega\hertz}$ ($B_\mathrm{RF} = \SI{69.8}{\micro\tesla}$), which satisfied the RWA condition for  single-RF-photon resonances, $\Omega_\mathrm{RF} \ll 2\tilde{M}_x (2V)$.
    Under the continuous application of MW and the RF field, we measured the ESDR spectra while setting the amplitude of the RF field at $B_\mathrm{RF} = \SI{69.8}{\micro\tesla}$ and varying its frequency.
    The results are shown in Fig~\ref{subfigure 2a}.
    anticrossing structures were observed when we set $\omega_\mathrm{RF}/(2\pi)=\SI{9.09}{\mega\hertz}$ [dashed line in Fig.~\ref{subfigure 2a}].
    The corresponding ESDR spectra are shown in Fig.~\ref{subfigure 2b}.
    This result indicates the formation of  RF-dressed states.
    Importantly, we did not observe the anticrossing structures near the RF frequency of $\omega_\mathrm{RF}/(2\pi) = \SI{4.5}{\mega\hertz}$ under the weak RF field.
    Here, we considered the effect of a parallel DC magnetic field in the calculations, because an unintentional parallel DC magnetic field may have been applied in the actual experiment.

    Using Eq.~\eqref{eq:averaged transition probability in Floquet theory}, we numerically simulate the ESDR spectra without (with) a parallel DC bias magnetic field, $\omega_L$, as shown in Fig~\ref{subfigure 2c} [Fig.~\ref{subfigure 2d}].
    We then truncated the Floquet Hamiltonians in Eqs.~\eqref{eq:Floquet Hamiltonian for ESDR} and \eqref{eq:Floquet Hamiltonian for ESDR with bias magnetic field} to $405 \times 405$ matrixes and set $\tilde{M}_x/(2\pi) = \SI{4.40}{\mega\hertz}$ and $\omega_L/(2\pi) = \SI{0.60}{\mega\hertz}$.
    Here, the parallel DC magnetic field, $B^{(\mathrm{bias})}_z = \omega_L/\gamma_e \approx \SI{0.02}{\milli\tesla}$, is relatively small compared with the Earth's magnetic field of $\approx \SI{0.05}{\milli\tesla}$.
    This is because the effect of the Earth's magnetic field and the misalignment of the perpendicular DC magnetic field probably cancel each other.
    Moreover, we extracted the resonant MW frequencies from Fig.~\ref{subfigure 2a} and compared them with the analytical solutions obtained using Eqs.~\eqref{eq:resonant MW frequencies of single-RF-photon resonance} and \eqref{eq:resonant MW frequencies of multi-RF-photon resonance}, as shown in Fig.~\ref{subfigure 2b}.

    With respect to the resonant MW frequencies, both the numerical and analytical solutions agreed with the experimental result.
    However, there was a small deviation between the experimental and theoretical results in the case of the contrast and linewidth.
    This is because we did not consider the effect of the initialization by the laser and the decoherence for simplicity.
    More importantly, we did not observe any significant differences between the numerical simulations performed with and without the parallel bias magnetic field, $\omega_L$, with respect to the resonant frequencies in the weak RF regime, $\Omega_\mathrm{RF} \ll 2\tilde{M}_x(2V)$.
    This is because the parallel magnetic field induces  multi-RF-photon resonances but does not significantly affect  single-RF-photon resonances.

\subsection{Strong RF Regime}
    \begin{figure}
        \centering
        \includegraphics[width=8.6cm]{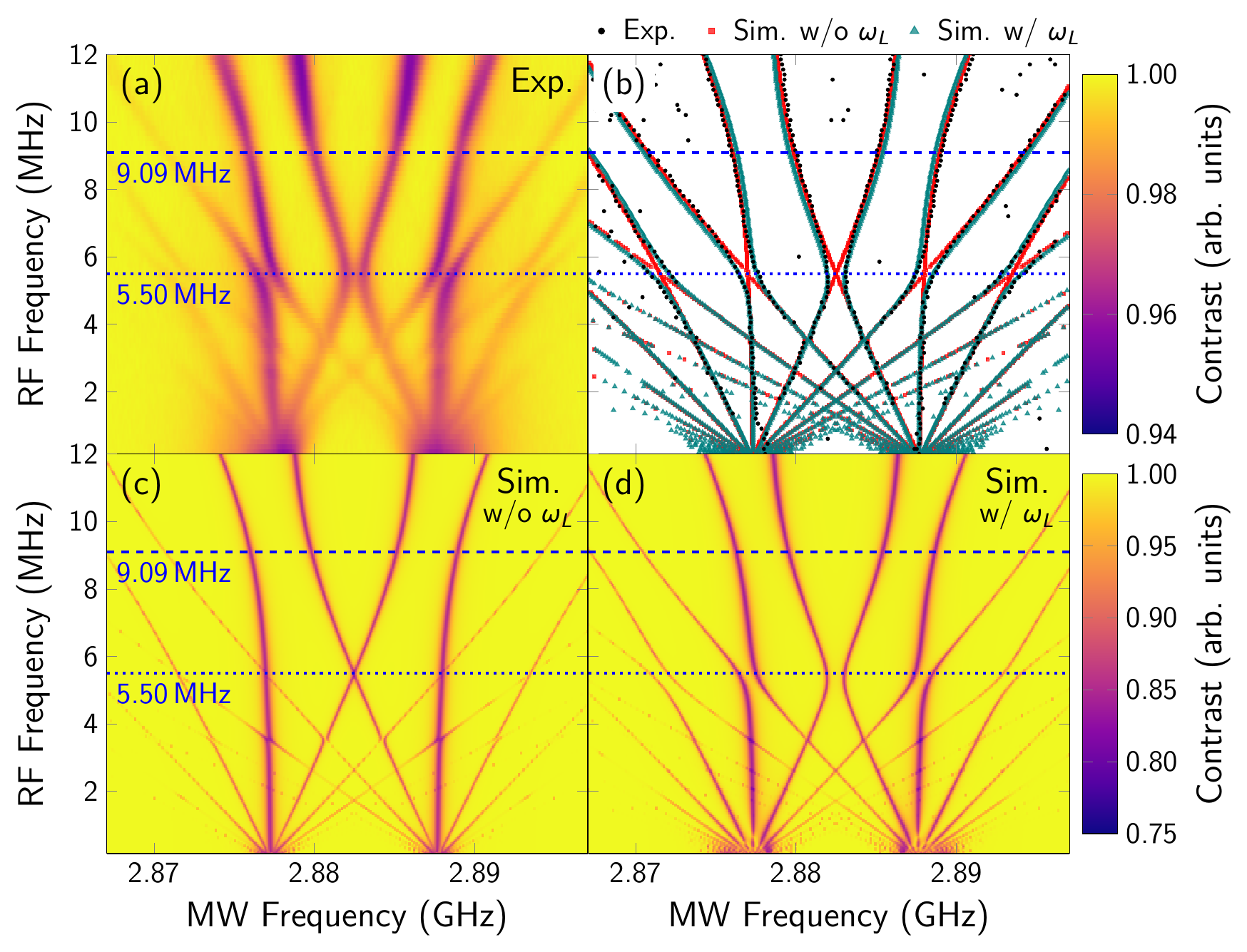}
        \phantomsubfloat{\label{subfigure 3a}}
        \phantomsubfloat{\label{subfigure 3b}}
        \phantomsubfloat{\label{subfigure 3c}}
        \phantomsubfloat{\label{subfigure 3d}}
        \caption{%
            (a) ESDR spectra under DC bias magnetic field perpendicular to NV axis as measured by sweeping MW frequencies and RF frequencies under strong RF field with $B_\mathrm{RF} = \SI{136.8}{\micro\tesla}$.
            (b) Comparison of extracted resonant frequencies obtained experimentally and those obtained from numerical simulations (c) and (d) under strong RF field.
            Experimental resonant frequencies were extracted from (a), and numerically determined were extracted from (c) and (d).
            (c) and (d) Results of numerical simulations performed using  Floquet Hamiltonian (c) without and (d) with parallel bias magnetic field.
            Parameters used are same as in Fig.~\ref{figure 2}.
            In (a)–(d), dashed lines indicate resonant frequency, $\omega_\mathrm{RF}/(2\pi) = \SI{9.09}{\mega\hertz}$, of  single-RF-photon resonances, and dotted lines indicate resonant frequency, $\omega_\mathrm{RF}/(2\pi) = \SI{5.50}{\mega\hertz}$, of two-RF-photon resonances. There is good agreement between experimental results and those of theoretical calculations that consider parallel bias magnetic field.
            In contrast, theory without parallel bias magnetic field could not reproduce anticrossing structures induced by two-RF-photon resonances.
        }
        \label{figure 3}
    \end{figure}
    In the second experiment, we applied a strong RF field with an amplitude of $B_\mathrm{RF} = \SI{136.8}{\micro\tesla}$ and measured the ESDR spectra while varying the frequency of the RF field.
    The RWA condition began to collapse  owing to the large RF amplitude, $\Omega_\mathrm{RF}$.
    Figure~\ref{subfigure 3a} shows the experimental results under the strong RF field.
    Similar to the case for the ESDR spectra under the weak RF field, the anticrossing structures corresponding to the single-RF-photon resonances were observed around the RF frequency, $\omega_\mathrm{RF}/(2\pi) = \SI{9.09}{\mega\hertz}$, with large splitting energy as indicated by the dashed line in Fig.~\ref{subfigure 3a}.
    In contrast to the weak RF regime, we observed  additional anticrossing structures when the RF frequency, $\omega_\mathrm{RF}/(2\pi)$, was approximately \SI{5.5}{\mega\hertz}, as indicated by the dotted line in Fig.~\ref{subfigure 3a}.
    The resonant RF frequency, $\omega_\mathrm{RF}/(2\pi)$, corresponding to these additional anticrossing structures were slightly different from the half of the energy gap between $\ket{B'}$ and $\ket{D'}$, that is, $V/(2\pi) = \SI{4.54}{\mega\hertz}$.
    This was owing to the Bloch-Siegert shift~\cite{Bloch1940} because of the large anticrossing structures near the RF frequency $\omega_\mathrm{RF}/(2\pi) = \SI{9.09}{\mega\hertz}$.

    Figures~\ref{subfigure 3c} and \ref{subfigure 3d} show the results of the numerical simulations of the truncated Floquet Hamiltonians (dimensions of $405 \times 405$) in Eqs.~\eqref{eq:Floquet Hamiltonian for ESDR} and \eqref{eq:Floquet Hamiltonian for ESDR with bias magnetic field}, respectively.
    Both simulations in Figs.~\ref{subfigure 3c} and \ref{subfigure 3d} could reproduce the anticrossing structures around the RF frequency $\omega_\mathrm{RF}/(2\pi) = \SI{9.09}{\mega\hertz}$ as well as the sideband resonances.
    However, the anticrossing structures specific to the strong RF field would not be reproduced by the simulations without the parallel bias magnetic field, $\omega_L$.
    In contrast, the simulation with the parallel bias magnetic field, $\omega_L$, could reproduce the anticrossing structures around the RF frequency, $\omega_\mathrm{RF}/(2\pi) = \SI{5.5}{\mega\hertz}$.
    For comparison, we extracted the resonant frequencies from  Figs.~\ref{subfigure 3a} and \ref{subfigure 3c} and plotted them in Figs.~\ref{subfigure 3b} and \ref{subfigure 3d}, respectively.
    The numerical results obtained considering parallel bias magnetic field, $\omega_L$, shown in Fig.~\ref{subfigure 3d} agreed with the experimental results shown in Fig.~\ref{subfigure 3a}.
    Therefore, the anticrossing structures specific to the strong RF field are induced by the two-RF-photon resonances allowed by the parallel bias magnetic field, as discussed in Sec.~\ref{sec:multi RF-photon resonance}.

    In this study, we only demonstrated single-RF-photon and two-RF-photon resonances.
    However, in principle, it should be possible to observe the anticrossing structures owing to the more RF-photon resonances by using a stronger RF field.

\subsection{Validity for Analytical Solutions}
    \begin{figure}
        \centering
        \includegraphics[width=8.6cm]{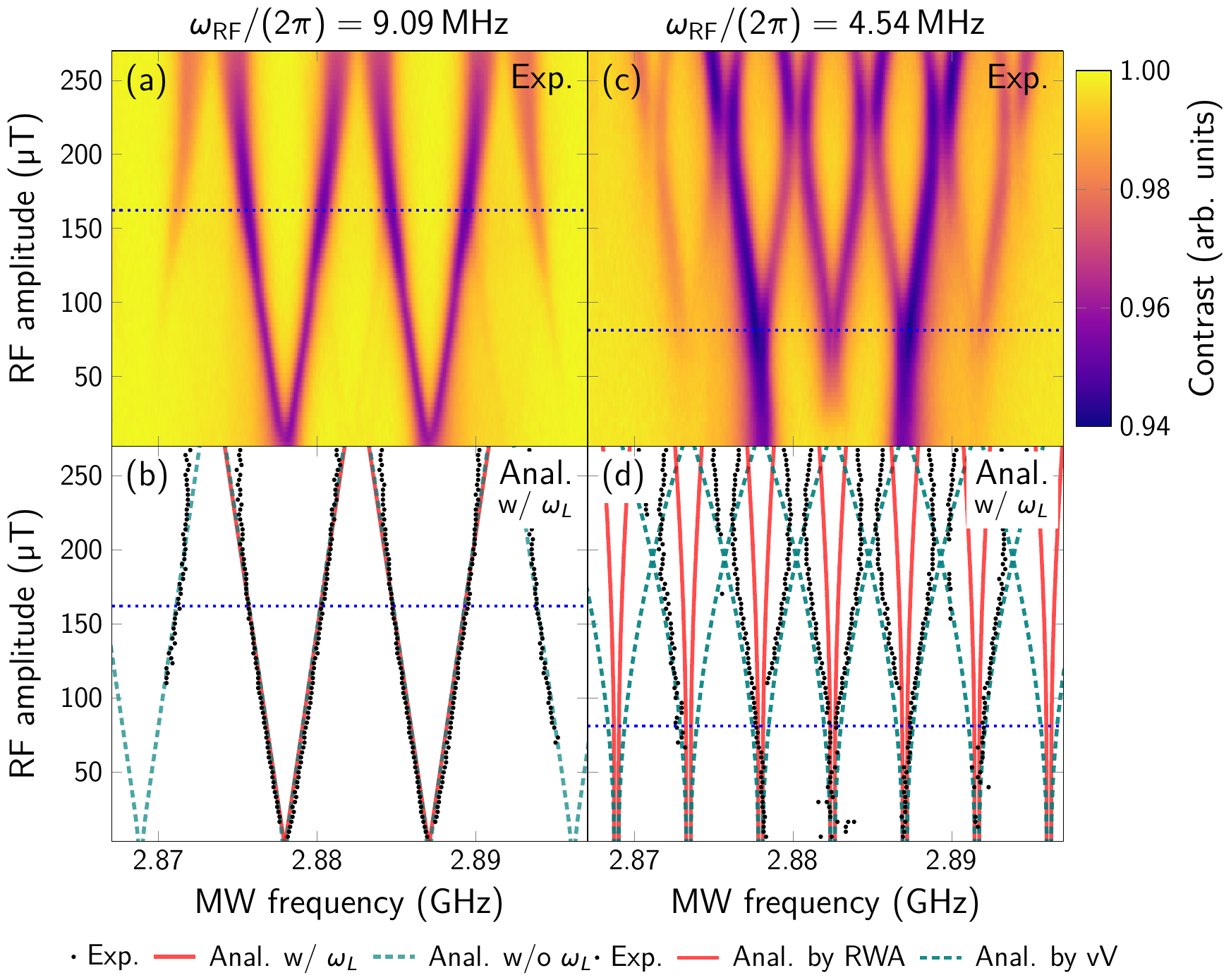}
        \phantomsubfloat{\label{subfigure 4a}}
        \phantomsubfloat{\label{subfigure 4b}}
        \phantomsubfloat{\label{subfigure 4c}}
        \phantomsubfloat{\label{subfigure 4d}}
        \caption{%
            (a) and (c) ESDR spectra under bias magnetic field perpendicular to NV axis as measured by changing RF amplitude $B_\mathrm{RF}$ while keeping the RF frequency at (a) $\omega_\mathrm{RF}/(2\pi) = \SI{9.09}{\mega\hertz}$ (single-RF-photon resonance) and (c) \SI{4.54}{\mega\hertz} (two-RF-photon resonance).
            (b) Comparison of resonant frequencies extracted from (a) and the analytical solutions given by Eqs.~\eqref{eq:resonant MW frequencies of single-RF-photon resonance} and \eqref{eq:resonant MW frequencies of multi-RF-photon resonance}.
            Analytical solutions obtained using Eqs.~\eqref{eq:resonant MW frequencies of single-RF-photon resonance} and \eqref{eq:resonant MW frequencies of multi-RF-photon resonance} which were obtained with and without a parallel DC magnetic field, respectively, agree with experimental results.
            (d) Comparison of resonant frequencies extracted from (c) and those obtained from analytical solutions given by Eq.~\eqref{eq:resonant MW frequencies of multi-RF-photon resonance}, which was based on RWA, and in Eq.~\eqref{eq:resonant MW frequencies of multi-RF-photon resonance by van Vleck transformation}, which was based on vV transformation.
            In (a) and (b) [(c) and (d)], dotted lines indicate boundaries of RWA conditions $\Omega_\mathrm{RF} > \omega_\mathrm{RF}/2$~\cite{Scheuer2014, Wang2021a}, which was calculated from $\omega_\mathrm{RF}/(2\pi) = \SI{9.09}{\mega\hertz}$, [$\SI{4.54}{MHz}$].
            Analytical solutions given by Eqs.~\eqref{eq:resonant MW frequencies of single-RF-photon resonance} and \eqref{eq:resonant MW frequencies of multi-RF-photon resonance} use RWA, which starts being violated in area above dotted line.
            In contrast, analytical solution given by Eq.~\eqref{eq:resonant MW frequencies of multi-RF-photon resonance by van Vleck transformation} does not use RWA and can reproduce resonant frequencies beyond RWA conditions.
        }
        \label{figure 4}
    \end{figure}
    In the third experiment, we fixed the RF frequency, $\omega_\mathrm{RF}/(2\pi)$, to $\SI{9.09}{\mega\hertz}$ while changing the RF amplitude, $B_\mathrm{RF}$, and measured the ESDR, as shown in Fig.~\ref{subfigure 4a}.
    As discussed in Refs.~\cite{Saijo2018, Dmitriev2019}, the anticrossing gaps induced by  single-RF-photon resonances increase linearly with an increase in the RF amplitude, $B_\mathrm{RF}$.
    We adopted the analytical solutions described by Eqs.~\eqref{eq:resonant MW frequencies of single-RF-photon resonance} and \eqref{eq:resonant MW frequencies of multi-RF-photon resonance} and attempted to fit the experimental results using them.
    In both cases, the calculation results agreed with the experimental ones.
    Moreover, the analytical solution in Eq.\eqref{eq:resonant MW frequencies of multi-RF-photon resonance} reproduced even the resonant frequencies of the sidebands.
    However, because we used the RWA to derive the analytical solutions in Eqs.~\eqref{eq:resonant MW frequencies of single-RF-photon resonance} and \eqref{eq:resonant MW frequencies of multi-RF-photon resonance}, these solutions will not be valid for strong RF fields.
    When the Rabi frequency of the RF field is larger than half the resonant RF frequency ($\Omega_\mathrm{RF} > \omega_\mathrm{RF}/2$)~\cite{Scheuer2014, Wang2021a}, the RWA is usually invalid.
    The analytical solutions in Eqs.~\eqref{eq:resonant MW frequencies of single-RF-photon resonance} and \eqref{eq:resonant MW frequencies of multi-RF-photon resonance} started to deviate from the experimental results at frequencies higher than the Rabi frequency, $\Omega_\mathrm{RF} > \SI{4.54}{\mega\hertz}$ (this RF amplitude is indicated by the dotted line in Figs.~\ref{subfigure 4a} and \ref{subfigure 4b}).

    To observe the two-RF-photon resonances, we measured the ESDR by setting the RF frequency as $\SI{4.54}{\mega\hertz}$ while changing the RF amplitude.
    The results are shown in Fig.~\ref{subfigure 4c}.
    The resonant RF frequencies of the two-RF-photon resonances were smaller than those of the single-RF-photon resonances.
    This implies that the RWA is violated for smaller RF amplitudes.
    Thus, we used the vV transformation to obtain the analytical solutions that would be valid for a strong RF field.
    In Fig.~\ref{subfigure 4d}, the experimentally measured resonant frequencies are plotted.
    In addition, we compared the results with the resonant frequencies calculated using the RWA and the vV transformation (See Eqs.~\eqref{eq:resonant MW frequencies of multi-RF-photon resonance} and \eqref{eq:resonant MW frequencies of multi-RF-photon resonance by van Vleck transformation}).
    When we used the analytical solutions with the RWA, we could not reproduce the experimental results for a strong RF field as shown in Fig.~\ref{subfigure 4d}.
    This is because the RWA is violated for $\Omega_\mathrm{RF} > \SI{2.27}{\mega\hertz}$ (indicated by the dotted line in Fig.~\ref{subfigure 4d}).
    However, the results obtained using the vV transformation (indicated by the dashed line) were in better agreement with the experimental results, as shown in Fig.~\ref{subfigure 4d}.
    This is because we considered a higher-order correction, as mentioned in Sec.\ref{sec:Analytical Solutions}.

\section{Conclusions and Future prospects}
    In this study, we theoretically and experimentally investigated the phenomenon of ESDR under strong RF fields.
    We observed the anticrossing structures attributable to multi-RF-photon resonances and reproduced the ESDR spectra of the NV centers through numerical simulations based on the Floquet theory.
    Moreover, using the vV transformation, we derived analytical solutions for the anticrossing structures observed under the strong RF field.
    Our results provide new insights into the phenomenon of ESDR, including the mechanism responsible for the anticrossing structures and the role of a parallel magnetic field.
    In addition, they should aid in the realization of practical RF sensors with NV centers and allow for the exploration of Floquet engineering in open quantum systems.

    Lastly, we discuss the direction for future work on the topic.
    Firstly, these results will aid the realization of practical RF sensors based on the CW-ODMR of  NV centers.
    Our understanding of the effect of a strong RF field on the phenomenon of ESDR.
    Because we elucidated the mechanism of the ESDR under  strong RF fields, it should be possible to develop practical RF sensors with a wider dynamic range.

    Next, the results of this study should help simulate various quantum phenomena in strong driving fields.
     Floquet engineering, which involves the creation of quantum systems with desired properties using a driving field has succeeded in realizing various quantum phenomena~\cite{Oka2019}.
    In such experiments, the quantum systems are well-designed isolated systems; the dissipation can be negligible~\cite{Ikeda2020, Ikeda2021, Mori2023}.
    However, most materials interact with the environment in reality, and their dissipation is not negligible.
    Moreover, it is usually difficult to detect the quantum states of such materials~\cite{Uchida2022}.
    Therefore, it is necessary to extend  Floquet engineering to open quantum systems.
    Because the NV center is robust to dissipation by the environment, and its spin state can be easily manipulated and readout with high fidelity, it is an ideal model for  Floquet engineering in open systems~\cite{Ikeda2020, Ikeda2021}.
    Strong driving (or multi-photon) phenomena in NV centers have been reported using different methods and setups~\cite{Fuchs2009, Childress2010, London2014, Tashima2019, Meinel2021, Nishimura2022}.
    However, in most previous studies, an MW field was used instead of an RF field.
    Because it is easier to go beyond the RWA regime in the case of RF fields compared with the MW, our approach may be more promising for observing phenomena owing to the breakdown of the approximation.
    Although a few studies have used an RF field, one needs to use multiple driving fields to realize strong driving phenomena; this results in high power consumption and requires complex control.
    Therefore, generating RF-dressed states using their approaches is not straightforward.

\begin{acknowledgments}
    We thank Dr. Kento Sasaki, Prof. Kensuke Kobayashi, and Dr. Ikeda N. Tatsuhiko for their valuable discussions.
    This work is supported by MEXT KAKENHI(20H05661, 22H01558), MEXT Q-LEAP(No. JPMXS0118067395),
    Leading Initiative for Excellent Young Researchers MEXT Japan, JST presto (Grant No. JPMJPR1919) Japan, JST (Moonshot R\&D)(Grant Number JPMJMS226C), and Kanazawa University CHOZEN Project 2022.
\end{acknowledgments}

\appendix

\section{Floquet Theory}\label{app:Floquet theory}
    Let us consider the time-dependent Schr\"{o}dinger equation with a time-periodic Hamiltonian ($\hbar=1$):
    \begin{equation}
        \hat{H}(t)\ket{\psi(t)} = i\diff*{\ket{\psi(t)}}{t}, \quad \hat{H}(t) = \hat{H}(t+T)
        \label{eq:Schrodinger equation with time-periodic Hamiltonian}
    \end{equation}
    where $T=\omega/(2\pi)$ denotes the time periodicity.
    Using Floquet's theorem, which is similar to Bloch's theorem, the solution of Eq.~\eqref{eq:Schrodinger equation with time-periodic Hamiltonian} can be written as a linear combination of the Floquet states, as follows:
    \begin{equation}
        \ket{\psi_\alpha(t)} =  e^{-i\varepsilon_\alpha t}\ket{\phi_\alpha(t)},
    \end{equation}
    Here, $\ket{\phi_\alpha(t)} = \ket{\phi_\alpha(t+T)}$ is the Floquet mode and $\varepsilon_\alpha$ is called the quasi-energy.
    We use the Fourier expansion,
    \begin{equation}
        \hat{H}(t) = \sum_m \hat{H}^{(m)}e^{i m \omega t}, \ \ket{\phi_\alpha(t)} = \sum_m e^{i m \omega t}\ket{\phi_\alpha^{(m)}}
    \end{equation}
    where $\hat{H}^{(n)}$ and $\ket{\phi_\alpha^{(n)}}$ are the $n$-th Fourier coefficients of $\hat{H}(t)$ and $\ket{\phi_\alpha^{(m)}}$, respectively.
    On substituting these into Eq.~\eqref{eq:Schrodinger equation with time-periodic Hamiltonian}, we obtain the infinite-dimensional time-independent eigenvalue equation~\cite{Shirley1965}
    \begin{equation}
        \sum_m \left[\hat{H}^{(n-m)} + m\omega\delta_{m,n}\hat{1}\right]\ket{\phi_\alpha^{(m)}} = \varepsilon_\alpha \ket{\phi_\alpha^{(n)}}
        \label{eq:Floquet-Schrodinger equation}
    \end{equation}
    where $\delta_{m,n}$ is the Kronecker delta and $\hat{1}$ is the identity operator.
    For convenience, we introduce the extended Hilbert space or the Sambe space, $\mathfrak{F} \coloneqq \mathfrak{H} \otimes \mathfrak{T}$~\cite{Sambe1973}, where $\mathfrak{H}$ and  $\mathfrak{T}$ are the Hilbert spaces for the quantum states and $T$-periodic functions, respectively.
    A quantum state $e^{i n \omega t}\ket{\alpha}$ in the Hilbert space corresponds to $\ket{\alpha, n} \coloneqq \ket{\alpha} \otimes \ket{n}$ in the Hilbert space.
    Moreover, we use the Floquet ladder operators $\hat{F}_n$ and the Floquet number operator, $\hat{N}$, as defined  in~\cite{Ivanov2021}:
    \begin{equation}
        \hat{\mathcal{F}}_n\ket{m} = \ket{n+m}, \quad \hat{\mathcal{N}}\ket{n} = n\ket{n}.
    \end{equation}
    Then, Eq.~\eqref{eq:Floquet-Schrodinger equation} can be simply written as
    \begin{equation}
        \hat{H}_F \ket{\phi_\alpha} = \varepsilon_\alpha \ket{\phi_\alpha}
        \label{eq:eigenvalue equation of Floquet Hamiltonian}
    \end{equation}
    where $\hat{H}_F$ is the Floquet Hamiltonian defined as
    \begin{align}
        \hat{H}_F &\coloneqq \sum_m \hat{\mathcal{F}}_m \otimes \hat{H}^{m} + \hat{\mathcal{N}} \otimes \omega\hat{1} \\
                   &=
                        \begin{bmatrix}
                            \ddots & \vdots & \vdots & \vdots &  \\
                            \cdots & \hat{H}^{(0)} + \omega\hat{1} & \hat{H}^{(+1)}  & \hat{H}^{(+2)} & \cdots \\
                            \cdots & \hat{H}^{(-1)} & \hat{H}^{(0)} & \hat{H}^{(+1)} & \cdots \\
                            \cdots & \hat{H}^{(-2)} & \hat{H}^{(-1)} & \hat{H}^{0} - \omega\hat{1} & \cdots \\
                             & \vdots & \vdots & \vdots & \ddots
                        \end{bmatrix}.
        \label{eq:Floquet-Schrodinger equation in Sambe space}
    \end{align}
    Therefore, the Floquet theory can be used to convert the time-dependent Schr\"{o}dinger equation given in Eq.~\eqref{eq:Schrodinger equation with time-periodic Hamiltonian} into an eigenvalue problem of the infinite-dimensional, time-independent Hamiltonian given in Eq.~\eqref{eq:Floquet-Schrodinger equation in Sambe space}.


\begin{thebibliography}{58}%
\makeatletter
\providecommand \@ifxundefined [1]{%
 \@ifx{#1\undefined}
}%
\providecommand \@ifnum [1]{%
 \ifnum #1\expandafter \@firstoftwo
 \else \expandafter \@secondoftwo
 \fi
}%
\providecommand \@ifx [1]{%
 \ifx #1\expandafter \@firstoftwo
 \else \expandafter \@secondoftwo
 \fi
}%
\providecommand \natexlab [1]{#1}%
\providecommand \enquote  [1]{``#1''}%
\providecommand \bibnamefont  [1]{#1}%
\providecommand \bibfnamefont [1]{#1}%
\providecommand \citenamefont [1]{#1}%
\providecommand \href@noop [0]{\@secondoftwo}%
\providecommand \href [0]{\begingroup \@sanitize@url \@href}%
\providecommand \@href[1]{\@@startlink{#1}\@@href}%
\providecommand \@@href[1]{\endgroup#1\@@endlink}%
\providecommand \@sanitize@url [0]{\catcode `\\12\catcode `\$12\catcode
  `\&12\catcode `\#12\catcode `\^12\catcode `\_12\catcode `\%12\relax}%
\providecommand \@@startlink[1]{}%
\providecommand \@@endlink[0]{}%
\providecommand \url  [0]{\begingroup\@sanitize@url \@url }%
\providecommand \@url [1]{\endgroup\@href {#1}{\urlprefix }}%
\providecommand \urlprefix  [0]{URL }%
\providecommand \Eprint [0]{\href }%
\providecommand \doibase [0]{http://dx.doi.org/}%
\providecommand \selectlanguage [0]{\@gobble}%
\providecommand \bibinfo  [0]{\@secondoftwo}%
\providecommand \bibfield  [0]{\@secondoftwo}%
\providecommand \translation [1]{[#1]}%
\providecommand \BibitemOpen [0]{}%
\providecommand \bibitemStop [0]{}%
\providecommand \bibitemNoStop [0]{.\EOS\space}%
\providecommand \EOS [0]{\spacefactor3000\relax}%
\providecommand \BibitemShut  [1]{\csname bibitem#1\endcsname}%
\let\auto@bib@innerbib\@empty
\bibitem [{\citenamefont {Levine}\ \emph {et~al.}(2019)\citenamefont {Levine},
  \citenamefont {Turner}, \citenamefont {Kehayias}, \citenamefont {Hart},
  \citenamefont {Langellier}, \citenamefont {Trubko}, \citenamefont {Glenn},
  \citenamefont {Fu},\ and\ \citenamefont {Walsworth}}]{Levine2019}%
  \BibitemOpen
  \bibfield  {author} {\bibinfo {author} {\bibfnamefont {E.~V.}\ \bibnamefont
  {Levine}}, \bibinfo {author} {\bibfnamefont {M.~J.}\ \bibnamefont {Turner}},
  \bibinfo {author} {\bibfnamefont {P.}~\bibnamefont {Kehayias}}, \bibinfo
  {author} {\bibfnamefont {C.~A.}\ \bibnamefont {Hart}}, \bibinfo {author}
  {\bibfnamefont {N.}~\bibnamefont {Langellier}}, \bibinfo {author}
  {\bibfnamefont {R.}~\bibnamefont {Trubko}}, \bibinfo {author} {\bibfnamefont
  {D.~R.}\ \bibnamefont {Glenn}}, \bibinfo {author} {\bibfnamefont {R.~R.}\
  \bibnamefont {Fu}}, \ and\ \bibinfo {author} {\bibfnamefont {R.~L.}\
  \bibnamefont {Walsworth}},\ }\href {\doibase 10.1515/nanoph-2019-0209}
  {\bibfield  {journal} {\bibinfo  {journal} {Nanophotonics}\ }\textbf
  {\bibinfo {volume} {8}},\ \bibinfo {pages} {1945} (\bibinfo {year}
  {2019})}\BibitemShut {NoStop}%
\bibitem [{\citenamefont {Barry}\ \emph {et~al.}(2020)\citenamefont {Barry},
  \citenamefont {Schloss}, \citenamefont {Bauch}, \citenamefont {Turner},
  \citenamefont {Hart}, \citenamefont {Pham},\ and\ \citenamefont
  {Walsworth}}]{Barry2020}%
  \BibitemOpen
  \bibfield  {author} {\bibinfo {author} {\bibfnamefont {J.~F.}\ \bibnamefont
  {Barry}}, \bibinfo {author} {\bibfnamefont {J.~M.}\ \bibnamefont {Schloss}},
  \bibinfo {author} {\bibfnamefont {E.}~\bibnamefont {Bauch}}, \bibinfo
  {author} {\bibfnamefont {M.~J.}\ \bibnamefont {Turner}}, \bibinfo {author}
  {\bibfnamefont {C.~A.}\ \bibnamefont {Hart}}, \bibinfo {author}
  {\bibfnamefont {L.~M.}\ \bibnamefont {Pham}}, \ and\ \bibinfo {author}
  {\bibfnamefont {R.~L.}\ \bibnamefont {Walsworth}},\ }\href {\doibase
  10.1103/RevModPhys.92.015004} {\bibfield  {journal} {\bibinfo  {journal}
  {Rev. Mod. Phys.}\ }\textbf {\bibinfo {volume} {92}},\ \bibinfo {pages}
  {015004} (\bibinfo {year} {2020})}\BibitemShut {NoStop}%
\bibitem [{\citenamefont {Ku}\ \emph {et~al.}(2020)\citenamefont {Ku},
  \citenamefont {Zhou}, \citenamefont {Li}, \citenamefont {Shin}, \citenamefont
  {Shi}, \citenamefont {Burch}, \citenamefont {Anderson}, \citenamefont
  {Pierce}, \citenamefont {Xie}, \citenamefont {Hamo}, \citenamefont {Vool},
  \citenamefont {Zhang}, \citenamefont {Casola}, \citenamefont {Taniguchi},
  \citenamefont {Watanabe}, \citenamefont {Fogler}, \citenamefont {Kim},
  \citenamefont {Yacoby},\ and\ \citenamefont {Walsworth}}]{Ku2020}%
  \BibitemOpen
  \bibfield  {author} {\bibinfo {author} {\bibfnamefont {M.~J.}\ \bibnamefont
  {Ku}}, \bibinfo {author} {\bibfnamefont {T.~X.}\ \bibnamefont {Zhou}},
  \bibinfo {author} {\bibfnamefont {Q.}~\bibnamefont {Li}}, \bibinfo {author}
  {\bibfnamefont {Y.~J.}\ \bibnamefont {Shin}}, \bibinfo {author}
  {\bibfnamefont {J.~K.}\ \bibnamefont {Shi}}, \bibinfo {author} {\bibfnamefont
  {C.}~\bibnamefont {Burch}}, \bibinfo {author} {\bibfnamefont {L.~E.}\
  \bibnamefont {Anderson}}, \bibinfo {author} {\bibfnamefont {A.~T.}\
  \bibnamefont {Pierce}}, \bibinfo {author} {\bibfnamefont {Y.}~\bibnamefont
  {Xie}}, \bibinfo {author} {\bibfnamefont {A.}~\bibnamefont {Hamo}}, \bibinfo
  {author} {\bibfnamefont {U.}~\bibnamefont {Vool}}, \bibinfo {author}
  {\bibfnamefont {H.}~\bibnamefont {Zhang}}, \bibinfo {author} {\bibfnamefont
  {F.}~\bibnamefont {Casola}}, \bibinfo {author} {\bibfnamefont
  {T.}~\bibnamefont {Taniguchi}}, \bibinfo {author} {\bibfnamefont
  {K.}~\bibnamefont {Watanabe}}, \bibinfo {author} {\bibfnamefont {M.~M.}\
  \bibnamefont {Fogler}}, \bibinfo {author} {\bibfnamefont {P.}~\bibnamefont
  {Kim}}, \bibinfo {author} {\bibfnamefont {A.}~\bibnamefont {Yacoby}}, \ and\
  \bibinfo {author} {\bibfnamefont {R.~L.}\ \bibnamefont {Walsworth}},\ }\href
  {\doibase 10.1038/s41586-020-2507-2} {\bibfield  {journal} {\bibinfo
  {journal} {Nature}\ }\textbf {\bibinfo {volume} {583}},\ \bibinfo {pages}
  {537} (\bibinfo {year} {2020})}\BibitemShut {NoStop}%
\bibitem [{\citenamefont {Huxter}\ \emph {et~al.}(2022)\citenamefont {Huxter},
  \citenamefont {Palm}, \citenamefont {Davis}, \citenamefont {Welter},
  \citenamefont {Lambert}, \citenamefont {Trassin},\ and\ \citenamefont
  {Degen}}]{Huxter2022}%
  \BibitemOpen
  \bibfield  {author} {\bibinfo {author} {\bibfnamefont {W.~S.}\ \bibnamefont
  {Huxter}}, \bibinfo {author} {\bibfnamefont {M.~L.}\ \bibnamefont {Palm}},
  \bibinfo {author} {\bibfnamefont {M.~L.}\ \bibnamefont {Davis}}, \bibinfo
  {author} {\bibfnamefont {P.}~\bibnamefont {Welter}}, \bibinfo {author}
  {\bibfnamefont {C.-H.}\ \bibnamefont {Lambert}}, \bibinfo {author}
  {\bibfnamefont {M.}~\bibnamefont {Trassin}}, \ and\ \bibinfo {author}
  {\bibfnamefont {C.~L.}\ \bibnamefont {Degen}},\ }\href {\doibase
  10.1038/s41467-022-31454-6} {\bibfield  {journal} {\bibinfo  {journal} {Nat.
  Commun.}\ }\textbf {\bibinfo {volume} {13}},\ \bibinfo {pages} {3761}
  (\bibinfo {year} {2022})}\BibitemShut {NoStop}%
\bibitem [{\citenamefont {Huxter}\ \emph {et~al.}(2023)\citenamefont {Huxter},
  \citenamefont {Sarott}, \citenamefont {Trassin},\ and\ \citenamefont
  {Degen}}]{Huxter2023}%
  \BibitemOpen
  \bibfield  {author} {\bibinfo {author} {\bibfnamefont {W.~S.}\ \bibnamefont
  {Huxter}}, \bibinfo {author} {\bibfnamefont {M.~F.}\ \bibnamefont {Sarott}},
  \bibinfo {author} {\bibfnamefont {M.}~\bibnamefont {Trassin}}, \ and\
  \bibinfo {author} {\bibfnamefont {C.~L.}\ \bibnamefont {Degen}},\ }\href
  {\doibase 10.1038/s41567-022-01921-4} {\bibfield  {journal} {\bibinfo
  {journal} {Nat. Phys.}\ } (\bibinfo {year} {2023}),\
  10.1038/s41567-022-01921-4}\BibitemShut {NoStop}%
\bibitem [{\citenamefont {Boyers}\ \emph {et~al.}(2020)\citenamefont {Boyers},
  \citenamefont {Crowley}, \citenamefont {Chandran},\ and\ \citenamefont
  {Sushkov}}]{Boyers2020}%
  \BibitemOpen
  \bibfield  {author} {\bibinfo {author} {\bibfnamefont {E.}~\bibnamefont
  {Boyers}}, \bibinfo {author} {\bibfnamefont {P.~J.~D.}\ \bibnamefont
  {Crowley}}, \bibinfo {author} {\bibfnamefont {A.}~\bibnamefont {Chandran}}, \
  and\ \bibinfo {author} {\bibfnamefont {A.~O.}\ \bibnamefont {Sushkov}},\
  }\href {\doibase 10.1103/PhysRevLett.125.160505} {\bibfield  {journal}
  {\bibinfo  {journal} {Phys. Rev. Lett.}\ }\textbf {\bibinfo {volume} {125}},\
  \bibinfo {pages} {160505} (\bibinfo {year} {2020})}\BibitemShut {NoStop}%
\bibitem [{\citenamefont {Zhang}\ \emph {et~al.}(2021)\citenamefont {Zhang},
  \citenamefont {Ouyang}, \citenamefont {Huang}, \citenamefont {Wang},
  \citenamefont {Zhang}, \citenamefont {Yu}, \citenamefont {Chang},
  \citenamefont {Liu}, \citenamefont {Deng},\ and\ \citenamefont
  {Duan}}]{Zhang2021}%
  \BibitemOpen
  \bibfield  {author} {\bibinfo {author} {\bibfnamefont {W.}~\bibnamefont
  {Zhang}}, \bibinfo {author} {\bibfnamefont {X.}~\bibnamefont {Ouyang}},
  \bibinfo {author} {\bibfnamefont {X.}~\bibnamefont {Huang}}, \bibinfo
  {author} {\bibfnamefont {X.}~\bibnamefont {Wang}}, \bibinfo {author}
  {\bibfnamefont {H.}~\bibnamefont {Zhang}}, \bibinfo {author} {\bibfnamefont
  {Y.}~\bibnamefont {Yu}}, \bibinfo {author} {\bibfnamefont {X.}~\bibnamefont
  {Chang}}, \bibinfo {author} {\bibfnamefont {Y.}~\bibnamefont {Liu}}, \bibinfo
  {author} {\bibfnamefont {D.-L.}\ \bibnamefont {Deng}}, \ and\ \bibinfo
  {author} {\bibfnamefont {L.-M.}\ \bibnamefont {Duan}},\ }\href {\doibase
  10.1103/PhysRevLett.127.090501} {\bibfield  {journal} {\bibinfo  {journal}
  {Phys. Rev. Lett.}\ }\textbf {\bibinfo {volume} {127}},\ \bibinfo {pages}
  {90501} (\bibinfo {year} {2021})}\BibitemShut {NoStop}%
\bibitem [{\citenamefont {Yang}\ \emph {et~al.}(2022)\citenamefont {Yang},
  \citenamefont {Xu}, \citenamefont {Zhou}, \citenamefont {Zhao}, \citenamefont
  {Xie}, \citenamefont {Ding}, \citenamefont {Ma}, \citenamefont {Gong},
  \citenamefont {Shi},\ and\ \citenamefont {Du}}]{Yang2022}%
  \BibitemOpen
  \bibfield  {author} {\bibinfo {author} {\bibfnamefont {K.}~\bibnamefont
  {Yang}}, \bibinfo {author} {\bibfnamefont {S.}~\bibnamefont {Xu}}, \bibinfo
  {author} {\bibfnamefont {L.}~\bibnamefont {Zhou}}, \bibinfo {author}
  {\bibfnamefont {Z.}~\bibnamefont {Zhao}}, \bibinfo {author} {\bibfnamefont
  {T.}~\bibnamefont {Xie}}, \bibinfo {author} {\bibfnamefont {Z.}~\bibnamefont
  {Ding}}, \bibinfo {author} {\bibfnamefont {W.}~\bibnamefont {Ma}}, \bibinfo
  {author} {\bibfnamefont {J.}~\bibnamefont {Gong}}, \bibinfo {author}
  {\bibfnamefont {F.}~\bibnamefont {Shi}}, \ and\ \bibinfo {author}
  {\bibfnamefont {J.}~\bibnamefont {Du}},\ }\href {\doibase
  10.1103/PhysRevB.106.184106} {\bibfield  {journal} {\bibinfo  {journal}
  {Phys. Rev. B}\ }\textbf {\bibinfo {volume} {106}},\ \bibinfo {pages}
  {184106} (\bibinfo {year} {2022})}\BibitemShut {NoStop}%
\bibitem [{\citenamefont {Schloss}\ \emph {et~al.}(2018)\citenamefont
  {Schloss}, \citenamefont {Barry}, \citenamefont {Turner},\ and\ \citenamefont
  {Walsworth}}]{Schloss2018}%
  \BibitemOpen
  \bibfield  {author} {\bibinfo {author} {\bibfnamefont {J.~M.}\ \bibnamefont
  {Schloss}}, \bibinfo {author} {\bibfnamefont {J.~F.}\ \bibnamefont {Barry}},
  \bibinfo {author} {\bibfnamefont {M.~J.}\ \bibnamefont {Turner}}, \ and\
  \bibinfo {author} {\bibfnamefont {R.~L.}\ \bibnamefont {Walsworth}},\ }\href
  {\doibase 10.1103/PhysRevApplied.10.034044} {\bibfield  {journal} {\bibinfo
  {journal} {Phys. Rev. Appl.}\ }\textbf {\bibinfo {volume} {10}},\ \bibinfo
  {pages} {34044} (\bibinfo {year} {2018})}\BibitemShut {NoStop}%
\bibitem [{\citenamefont {Tsukamoto}\ \emph {et~al.}(2021)\citenamefont
  {Tsukamoto}, \citenamefont {Ogawa}, \citenamefont {Ozawa}, \citenamefont
  {Iwasaki}, \citenamefont {Hatano}, \citenamefont {Sasaki},\ and\
  \citenamefont {Kobayashi}}]{Tsukamoto2021}%
  \BibitemOpen
  \bibfield  {author} {\bibinfo {author} {\bibfnamefont {M.}~\bibnamefont
  {Tsukamoto}}, \bibinfo {author} {\bibfnamefont {K.}~\bibnamefont {Ogawa}},
  \bibinfo {author} {\bibfnamefont {H.}~\bibnamefont {Ozawa}}, \bibinfo
  {author} {\bibfnamefont {T.}~\bibnamefont {Iwasaki}}, \bibinfo {author}
  {\bibfnamefont {M.}~\bibnamefont {Hatano}}, \bibinfo {author} {\bibfnamefont
  {K.}~\bibnamefont {Sasaki}}, \ and\ \bibinfo {author} {\bibfnamefont
  {K.}~\bibnamefont {Kobayashi}},\ }\href {\doibase 10.1063/5.0054809}
  {\bibfield  {journal} {\bibinfo  {journal} {Appl. Phys. Lett.}\ }\textbf
  {\bibinfo {volume} {118}},\ \bibinfo {pages} {264002} (\bibinfo {year}
  {2021})}\BibitemShut {NoStop}%
\bibitem [{\citenamefont {Chen}\ \emph {et~al.}(2022)\citenamefont {Chen},
  \citenamefont {Chen}, \citenamefont {Zhu}, \citenamefont {Fan}, \citenamefont
  {Yu}, \citenamefont {Qian},\ and\ \citenamefont {Xu}}]{Chen2022}%
  \BibitemOpen
  \bibfield  {author} {\bibinfo {author} {\bibfnamefont {B.}~\bibnamefont
  {Chen}}, \bibinfo {author} {\bibfnamefont {B.}~\bibnamefont {Chen}}, \bibinfo
  {author} {\bibfnamefont {X.}~\bibnamefont {Zhu}}, \bibinfo {author}
  {\bibfnamefont {J.}~\bibnamefont {Fan}}, \bibinfo {author} {\bibfnamefont
  {Z.}~\bibnamefont {Yu}}, \bibinfo {author} {\bibfnamefont {P.}~\bibnamefont
  {Qian}}, \ and\ \bibinfo {author} {\bibfnamefont {N.}~\bibnamefont {Xu}},\
  }\href {\doibase 10.1063/5.0121925} {\bibfield  {journal} {\bibinfo
  {journal} {Rev. Sci. Instrum.}\ }\textbf {\bibinfo {volume} {93}},\ \bibinfo
  {pages} {125105} (\bibinfo {year} {2022})}\BibitemShut {NoStop}%
\bibitem [{\citenamefont {Simon}\ \emph {et~al.}(2017)\citenamefont {Simon},
  \citenamefont {Tuvia}, \citenamefont {St{\"{u}}rner}, \citenamefont {Thomas},
  \citenamefont {Gerhard}, \citenamefont {Christoph}, \citenamefont {Jochen},
  \citenamefont {Boris}, \citenamefont {Matthew}, \citenamefont {Sebastien},
  \citenamefont {Jan}, \citenamefont {Ilai}, \citenamefont {Martin},
  \citenamefont {Alex}, \citenamefont {McGuinness},\ and\ \citenamefont
  {Fedor}}]{Simon2017}%
  \BibitemOpen
  \bibfield  {author} {\bibinfo {author} {\bibfnamefont {S.}~\bibnamefont
  {Simon}}, \bibinfo {author} {\bibfnamefont {G.}~\bibnamefont {Tuvia}},
  \bibinfo {author} {\bibfnamefont {F.~M.}\ \bibnamefont {St{\"{u}}rner}},
  \bibinfo {author} {\bibfnamefont {U.}~\bibnamefont {Thomas}}, \bibinfo
  {author} {\bibfnamefont {W.}~\bibnamefont {Gerhard}}, \bibinfo {author}
  {\bibfnamefont {M.}~\bibnamefont {Christoph}}, \bibinfo {author}
  {\bibfnamefont {S.}~\bibnamefont {Jochen}}, \bibinfo {author} {\bibfnamefont
  {N.}~\bibnamefont {Boris}}, \bibinfo {author} {\bibfnamefont
  {M.}~\bibnamefont {Matthew}}, \bibinfo {author} {\bibfnamefont
  {P.}~\bibnamefont {Sebastien}}, \bibinfo {author} {\bibfnamefont
  {M.}~\bibnamefont {Jan}}, \bibinfo {author} {\bibfnamefont {S.}~\bibnamefont
  {Ilai}}, \bibinfo {author} {\bibfnamefont {P.}~\bibnamefont {Martin}},
  \bibinfo {author} {\bibfnamefont {R.}~\bibnamefont {Alex}}, \bibinfo {author}
  {\bibfnamefont {L.~P.}\ \bibnamefont {McGuinness}}, \ and\ \bibinfo {author}
  {\bibfnamefont {J.}~\bibnamefont {Fedor}},\ }\href {\doibase
  10.1126/science.aam5532} {\bibfield  {journal} {\bibinfo  {journal}
  {Science}\ }\textbf {\bibinfo {volume} {356}},\ \bibinfo {pages} {832}
  (\bibinfo {year} {2017})}\BibitemShut {NoStop}%
\bibitem [{\citenamefont {Boss}\ \emph {et~al.}(2017)\citenamefont {Boss},
  \citenamefont {Cujia}, \citenamefont {Zopes},\ and\ \citenamefont
  {Degen}}]{Boss2017}%
  \BibitemOpen
  \bibfield  {author} {\bibinfo {author} {\bibfnamefont {J.~M.}\ \bibnamefont
  {Boss}}, \bibinfo {author} {\bibfnamefont {K.~S.}\ \bibnamefont {Cujia}},
  \bibinfo {author} {\bibfnamefont {J.}~\bibnamefont {Zopes}}, \ and\ \bibinfo
  {author} {\bibfnamefont {C.~L.}\ \bibnamefont {Degen}},\ }\href {\doibase
  10.1126/science.aam7009} {\bibfield  {journal} {\bibinfo  {journal}
  {Science}\ }\textbf {\bibinfo {volume} {356}},\ \bibinfo {pages} {837}
  (\bibinfo {year} {2017})}\BibitemShut {NoStop}%
\bibitem [{\citenamefont {Hart}\ \emph {et~al.}(2021)\citenamefont {Hart},
  \citenamefont {Schloss}, \citenamefont {Turner}, \citenamefont {Scheidegger},
  \citenamefont {Bauch},\ and\ \citenamefont {Walsworth}}]{Hart2021}%
  \BibitemOpen
  \bibfield  {author} {\bibinfo {author} {\bibfnamefont {C.~A.}\ \bibnamefont
  {Hart}}, \bibinfo {author} {\bibfnamefont {J.~M.}\ \bibnamefont {Schloss}},
  \bibinfo {author} {\bibfnamefont {M.~J.}\ \bibnamefont {Turner}}, \bibinfo
  {author} {\bibfnamefont {P.~J.}\ \bibnamefont {Scheidegger}}, \bibinfo
  {author} {\bibfnamefont {E.}~\bibnamefont {Bauch}}, \ and\ \bibinfo {author}
  {\bibfnamefont {R.~L.}\ \bibnamefont {Walsworth}},\ }\href {\doibase
  10.1103/PhysRevApplied.15.044020} {\bibfield  {journal} {\bibinfo  {journal}
  {Phys. Rev. Appl.}\ }\textbf {\bibinfo {volume} {15}},\ \bibinfo {pages}
  {44020} (\bibinfo {year} {2021})}\BibitemShut {NoStop}%
\bibitem [{\citenamefont {Wang}\ \emph {et~al.}(2022)\citenamefont {Wang},
  \citenamefont {Liu}, \citenamefont {Schloss}, \citenamefont {Alsid},
  \citenamefont {Braje},\ and\ \citenamefont {Cappellaro}}]{Wang2022}%
  \BibitemOpen
  \bibfield  {author} {\bibinfo {author} {\bibfnamefont {G.}~\bibnamefont
  {Wang}}, \bibinfo {author} {\bibfnamefont {Y.-X.}\ \bibnamefont {Liu}},
  \bibinfo {author} {\bibfnamefont {J.~M.}\ \bibnamefont {Schloss}}, \bibinfo
  {author} {\bibfnamefont {S.~T.}\ \bibnamefont {Alsid}}, \bibinfo {author}
  {\bibfnamefont {D.~A.}\ \bibnamefont {Braje}}, \ and\ \bibinfo {author}
  {\bibfnamefont {P.}~\bibnamefont {Cappellaro}},\ }\href {\doibase
  10.1103/PhysRevX.12.021061} {\bibfield  {journal} {\bibinfo  {journal} {Phys.
  Rev. X}\ }\textbf {\bibinfo {volume} {12}},\ \bibinfo {pages} {21061}
  (\bibinfo {year} {2022})}\BibitemShut {NoStop}%
\bibitem [{\citenamefont {Saijo}\ \emph {et~al.}(2018)\citenamefont {Saijo},
  \citenamefont {Matsuzaki}, \citenamefont {Saito}, \citenamefont {Yamaguchi},
  \citenamefont {Hanano}, \citenamefont {Watanabe}, \citenamefont {Mizuochi},\
  and\ \citenamefont {Ishi-Hayase}}]{Saijo2018}%
  \BibitemOpen
  \bibfield  {author} {\bibinfo {author} {\bibfnamefont {S.}~\bibnamefont
  {Saijo}}, \bibinfo {author} {\bibfnamefont {Y.}~\bibnamefont {Matsuzaki}},
  \bibinfo {author} {\bibfnamefont {S.}~\bibnamefont {Saito}}, \bibinfo
  {author} {\bibfnamefont {T.}~\bibnamefont {Yamaguchi}}, \bibinfo {author}
  {\bibfnamefont {I.}~\bibnamefont {Hanano}}, \bibinfo {author} {\bibfnamefont
  {H.}~\bibnamefont {Watanabe}}, \bibinfo {author} {\bibfnamefont
  {N.}~\bibnamefont {Mizuochi}}, \ and\ \bibinfo {author} {\bibfnamefont
  {J.}~\bibnamefont {Ishi-Hayase}},\ }\href {\doibase 10.1063/1.5024401}
  {\bibfield  {journal} {\bibinfo  {journal} {Appl. Phys. Lett.}\ }\textbf
  {\bibinfo {volume} {113}},\ \bibinfo {pages} {082405} (\bibinfo {year}
  {2018})}\BibitemShut {NoStop}%
\bibitem [{\citenamefont {Yamaguchi}\ \emph {et~al.}(2019)\citenamefont
  {Yamaguchi}, \citenamefont {Matsuzaki}, \citenamefont {Saito}, \citenamefont
  {Saijo}, \citenamefont {Watanabe}, \citenamefont {Mizuochi},\ and\
  \citenamefont {Ishi-Hayase}}]{Yamaguchi2019}%
  \BibitemOpen
  \bibfield  {author} {\bibinfo {author} {\bibfnamefont {T.}~\bibnamefont
  {Yamaguchi}}, \bibinfo {author} {\bibfnamefont {Y.}~\bibnamefont
  {Matsuzaki}}, \bibinfo {author} {\bibfnamefont {S.}~\bibnamefont {Saito}},
  \bibinfo {author} {\bibfnamefont {S.}~\bibnamefont {Saijo}}, \bibinfo
  {author} {\bibfnamefont {H.}~\bibnamefont {Watanabe}}, \bibinfo {author}
  {\bibfnamefont {N.}~\bibnamefont {Mizuochi}}, \ and\ \bibinfo {author}
  {\bibfnamefont {J.}~\bibnamefont {Ishi-Hayase}},\ }\href {\doibase
  10.7567/1347-4065/ab3d03} {\bibfield  {journal} {\bibinfo  {journal} {Jpn. J.
  Appl. Phys.}\ }\textbf {\bibinfo {volume} {58}},\ \bibinfo {pages} {100901}
  (\bibinfo {year} {2019})}\BibitemShut {NoStop}%
\bibitem [{\citenamefont {Tabuchi}\ \emph {et~al.}(2023)\citenamefont
  {Tabuchi}, \citenamefont {Matsuzaki}, \citenamefont {Furuya}, \citenamefont
  {Nakano}, \citenamefont {Watanabe}, \citenamefont {Tokuda}, \citenamefont
  {Mizuochi},\ and\ \citenamefont {Ishi-Hayase}}]{Tabuchi2023}%
  \BibitemOpen
  \bibfield  {author} {\bibinfo {author} {\bibfnamefont {H.}~\bibnamefont
  {Tabuchi}}, \bibinfo {author} {\bibfnamefont {Y.}~\bibnamefont {Matsuzaki}},
  \bibinfo {author} {\bibfnamefont {N.}~\bibnamefont {Furuya}}, \bibinfo
  {author} {\bibfnamefont {Y.}~\bibnamefont {Nakano}}, \bibinfo {author}
  {\bibfnamefont {H.}~\bibnamefont {Watanabe}}, \bibinfo {author}
  {\bibfnamefont {N.}~\bibnamefont {Tokuda}}, \bibinfo {author} {\bibfnamefont
  {N.}~\bibnamefont {Mizuochi}}, \ and\ \bibinfo {author} {\bibfnamefont
  {J.}~\bibnamefont {Ishi-Hayase}},\ }\href {\doibase 10.1063/5.0129706}
  {\bibfield  {journal} {\bibinfo  {journal} {J. Appl. Phys.}\ }\textbf
  {\bibinfo {volume} {133}},\ \bibinfo {pages} {24401} (\bibinfo {year}
  {2023})}\BibitemShut {NoStop}%
\bibitem [{\citenamefont {Dmitriev}\ and\ \citenamefont
  {Vershovskii}(2018)}]{Dmitriev2018}%
  \BibitemOpen
  \bibfield  {author} {\bibinfo {author} {\bibfnamefont {A.~K.}\ \bibnamefont
  {Dmitriev}}\ and\ \bibinfo {author} {\bibfnamefont {A.~K.}\ \bibnamefont
  {Vershovskii}},\ }\href {\doibase 10.1088/1742-6596/1135/1/012051} {\bibfield
   {journal} {\bibinfo  {journal} {J. Phys. Conf. Ser.}\ }\textbf {\bibinfo
  {volume} {1135}},\ \bibinfo {pages} {12051} (\bibinfo {year}
  {2018})}\BibitemShut {NoStop}%
\bibitem [{\citenamefont {Dmitriev}\ \emph {et~al.}(2019)\citenamefont
  {Dmitriev}, \citenamefont {Chen}, \citenamefont {Fuchs},\ and\ \citenamefont
  {Vershovskii}}]{Dmitriev2019}%
  \BibitemOpen
  \bibfield  {author} {\bibinfo {author} {\bibfnamefont {A.~K.}\ \bibnamefont
  {Dmitriev}}, \bibinfo {author} {\bibfnamefont {H.~Y.}\ \bibnamefont {Chen}},
  \bibinfo {author} {\bibfnamefont {G.~D.}\ \bibnamefont {Fuchs}}, \ and\
  \bibinfo {author} {\bibfnamefont {A.~K.}\ \bibnamefont {Vershovskii}},\
  }\href {\doibase 10.1103/PhysRevA.100.011801} {\bibfield  {journal} {\bibinfo
   {journal} {Phys. Rev. A}\ }\textbf {\bibinfo {volume} {100}},\ \bibinfo
  {pages} {11801} (\bibinfo {year} {2019})}\BibitemShut {NoStop}%
\bibitem [{\citenamefont {Dmitriev}\ and\ \citenamefont
  {Vershovskii}(2020)}]{Dmitriev2020}%
  \BibitemOpen
  \bibfield  {author} {\bibinfo {author} {\bibfnamefont {A.~K.}\ \bibnamefont
  {Dmitriev}}\ and\ \bibinfo {author} {\bibfnamefont {A.~K.}\ \bibnamefont
  {Vershovskii}},\ }\href {\doibase 10.1109i/LSENS.2019.2957328} {\bibfield
  {journal} {\bibinfo  {journal} {IEEE Sens. Lett.}\ }\textbf {\bibinfo
  {volume} {4}},\ \bibinfo {pages} {1} (\bibinfo {year} {2020})}\BibitemShut
  {NoStop}%
\bibitem [{\citenamefont {Dmitriev}\ and\ \citenamefont
  {Vershovskii}(2022)}]{Dmitriev2022}%
  \BibitemOpen
  \bibfield  {author} {\bibinfo {author} {\bibfnamefont {A.~K.}\ \bibnamefont
  {Dmitriev}}\ and\ \bibinfo {author} {\bibfnamefont {A.~K.}\ \bibnamefont
  {Vershovskii}},\ }\href {\doibase 10.1103/PhysRevA.105.043509} {\bibfield
  {journal} {\bibinfo  {journal} {Phys. Rev. A}\ }\textbf {\bibinfo {volume}
  {105}},\ \bibinfo {pages} {43509} (\bibinfo {year} {2022})}\BibitemShut
  {NoStop}%
\bibitem [{\citenamefont {Fuchs}\ \emph {et~al.}(2009)\citenamefont {Fuchs},
  \citenamefont {Dobrovitski}, \citenamefont {Toyli}, \citenamefont
  {Heremans},\ and\ \citenamefont {Awschalom}}]{Fuchs2009}%
  \BibitemOpen
  \bibfield  {author} {\bibinfo {author} {\bibfnamefont {G.~D.}\ \bibnamefont
  {Fuchs}}, \bibinfo {author} {\bibfnamefont {V.~V.}\ \bibnamefont
  {Dobrovitski}}, \bibinfo {author} {\bibfnamefont {D.~M.}\ \bibnamefont
  {Toyli}}, \bibinfo {author} {\bibfnamefont {F.~J.}\ \bibnamefont {Heremans}},
  \ and\ \bibinfo {author} {\bibfnamefont {D.~D.}\ \bibnamefont {Awschalom}},\
  }\href {\doibase 10.1126/science.1181193} {\bibfield  {journal} {\bibinfo
  {journal} {Science}\ }\textbf {\bibinfo {volume} {326}},\ \bibinfo {pages}
  {1520} (\bibinfo {year} {2009})}\BibitemShut {NoStop}%
\bibitem [{\citenamefont {London}\ \emph {et~al.}(2014)\citenamefont {London},
  \citenamefont {Balasubramanian}, \citenamefont {Naydenov}, \citenamefont
  {McGuinness},\ and\ \citenamefont {Jelezko}}]{London2014}%
  \BibitemOpen
  \bibfield  {author} {\bibinfo {author} {\bibfnamefont {P.}~\bibnamefont
  {London}}, \bibinfo {author} {\bibfnamefont {P.}~\bibnamefont
  {Balasubramanian}}, \bibinfo {author} {\bibfnamefont {B.}~\bibnamefont
  {Naydenov}}, \bibinfo {author} {\bibfnamefont {L.~P.}\ \bibnamefont
  {McGuinness}}, \ and\ \bibinfo {author} {\bibfnamefont {F.}~\bibnamefont
  {Jelezko}},\ }\href {\doibase 10.1103/PhysRevA.90.012302} {\bibfield
  {journal} {\bibinfo  {journal} {Phys. Rev. A}\ }\textbf {\bibinfo {volume}
  {90}},\ \bibinfo {pages} {12302} (\bibinfo {year} {2014})}\BibitemShut
  {NoStop}%
\bibitem [{\citenamefont {Scheuer}\ \emph {et~al.}(2014)\citenamefont
  {Scheuer}, \citenamefont {Kong}, \citenamefont {Said}, \citenamefont {Chen},
  \citenamefont {Kurz}, \citenamefont {Marseglia}, \citenamefont {Du},
  \citenamefont {Hemmer}, \citenamefont {Montangero}, \citenamefont {Calarco},
  \citenamefont {Naydenov},\ and\ \citenamefont {Jelezko}}]{Scheuer2014}%
  \BibitemOpen
  \bibfield  {author} {\bibinfo {author} {\bibfnamefont {J.}~\bibnamefont
  {Scheuer}}, \bibinfo {author} {\bibfnamefont {X.}~\bibnamefont {Kong}},
  \bibinfo {author} {\bibfnamefont {R.~S.}\ \bibnamefont {Said}}, \bibinfo
  {author} {\bibfnamefont {J.}~\bibnamefont {Chen}}, \bibinfo {author}
  {\bibfnamefont {A.}~\bibnamefont {Kurz}}, \bibinfo {author} {\bibfnamefont
  {L.}~\bibnamefont {Marseglia}}, \bibinfo {author} {\bibfnamefont
  {J.}~\bibnamefont {Du}}, \bibinfo {author} {\bibfnamefont {P.~R.}\
  \bibnamefont {Hemmer}}, \bibinfo {author} {\bibfnamefont {S.}~\bibnamefont
  {Montangero}}, \bibinfo {author} {\bibfnamefont {T.}~\bibnamefont {Calarco}},
  \bibinfo {author} {\bibfnamefont {B.}~\bibnamefont {Naydenov}}, \ and\
  \bibinfo {author} {\bibfnamefont {F.}~\bibnamefont {Jelezko}},\ }\href
  {\doibase 10.1088/1367-2630/16/9/093022} {\bibfield  {journal} {\bibinfo
  {journal} {New J. Phys.}\ }\textbf {\bibinfo {volume} {16}},\ \bibinfo
  {pages} {093022} (\bibinfo {year} {2014})}\BibitemShut {NoStop}%
\bibitem [{\citenamefont {Laucht}\ \emph {et~al.}(2016)\citenamefont {Laucht},
  \citenamefont {Simmons}, \citenamefont {Kalra}, \citenamefont {Tosi},
  \citenamefont {Dehollain}, \citenamefont {Muhonen}, \citenamefont {Freer},
  \citenamefont {Hudson}, \citenamefont {Itoh}, \citenamefont {Jamieson},
  \citenamefont {McCallum}, \citenamefont {Dzurak},\ and\ \citenamefont
  {Morello}}]{Laucht2016}%
  \BibitemOpen
  \bibfield  {author} {\bibinfo {author} {\bibfnamefont {A.}~\bibnamefont
  {Laucht}}, \bibinfo {author} {\bibfnamefont {S.}~\bibnamefont {Simmons}},
  \bibinfo {author} {\bibfnamefont {R.}~\bibnamefont {Kalra}}, \bibinfo
  {author} {\bibfnamefont {G.}~\bibnamefont {Tosi}}, \bibinfo {author}
  {\bibfnamefont {J.~P.}\ \bibnamefont {Dehollain}}, \bibinfo {author}
  {\bibfnamefont {J.~T.}\ \bibnamefont {Muhonen}}, \bibinfo {author}
  {\bibfnamefont {S.}~\bibnamefont {Freer}}, \bibinfo {author} {\bibfnamefont
  {F.~E.}\ \bibnamefont {Hudson}}, \bibinfo {author} {\bibfnamefont {K.~M.}\
  \bibnamefont {Itoh}}, \bibinfo {author} {\bibfnamefont {D.~N.}\ \bibnamefont
  {Jamieson}}, \bibinfo {author} {\bibfnamefont {J.~C.}\ \bibnamefont
  {McCallum}}, \bibinfo {author} {\bibfnamefont {A.~S.}\ \bibnamefont
  {Dzurak}}, \ and\ \bibinfo {author} {\bibfnamefont {A.}~\bibnamefont
  {Morello}},\ }\href {\doibase 10.1103/PhysRevB.94.161302} {\bibfield
  {journal} {\bibinfo  {journal} {Phys. Rev. B}\ }\textbf {\bibinfo {volume}
  {94}},\ \bibinfo {pages} {161302} (\bibinfo {year} {2016})}\BibitemShut
  {NoStop}%
\bibitem [{\citenamefont {Shirley}(1965)}]{Shirley1965}%
  \BibitemOpen
  \bibfield  {author} {\bibinfo {author} {\bibfnamefont {J.~H.}\ \bibnamefont
  {Shirley}},\ }\href {\doibase 10.1103/PhysRev.138.B979} {\bibfield  {journal}
  {\bibinfo  {journal} {Phys. Rev.}\ }\textbf {\bibinfo {volume} {138}},\
  \bibinfo {pages} {B979} (\bibinfo {year} {1965})}\BibitemShut {NoStop}%
\bibitem [{\citenamefont {Sambe}(1973)}]{Sambe1973}%
  \BibitemOpen
  \bibfield  {author} {\bibinfo {author} {\bibfnamefont {H.}~\bibnamefont
  {Sambe}},\ }\href {\doibase 10.1103/PhysRevA.7.2203} {\bibfield  {journal}
  {\bibinfo  {journal} {Phys. Rev. A}\ }\textbf {\bibinfo {volume} {7}},\
  \bibinfo {pages} {2203} (\bibinfo {year} {1973})}\BibitemShut {NoStop}%
\bibitem [{\citenamefont {Eckardt}\ and\ \citenamefont
  {Anisimovas}(2015)}]{Eckardt2015}%
  \BibitemOpen
  \bibfield  {author} {\bibinfo {author} {\bibfnamefont {A.}~\bibnamefont
  {Eckardt}}\ and\ \bibinfo {author} {\bibfnamefont {E.}~\bibnamefont
  {Anisimovas}},\ }\href {\doibase 10.1088/1367-2630/17/9/093039} {\bibfield
  {journal} {\bibinfo  {journal} {New J. Phys.}\ }\textbf {\bibinfo {volume}
  {17}},\ \bibinfo {pages} {093039} (\bibinfo {year} {2015})}\BibitemShut
  {NoStop}%
\bibitem [{\citenamefont {Oka}\ and\ \citenamefont {Kitamura}(2019)}]{Oka2019}%
  \BibitemOpen
  \bibfield  {author} {\bibinfo {author} {\bibfnamefont {T.}~\bibnamefont
  {Oka}}\ and\ \bibinfo {author} {\bibfnamefont {S.}~\bibnamefont {Kitamura}},\
  }\href {\doibase 10.1146/annurev-conmatphys-031218-013423} {\bibfield
  {journal} {\bibinfo  {journal} {Annu. Rev. Condens. Matter Phys.}\ }\textbf
  {\bibinfo {volume} {10}},\ \bibinfo {pages} {387} (\bibinfo {year}
  {2019})}\BibitemShut {NoStop}%
\bibitem [{\citenamefont {Ivanov}\ \emph {et~al.}(2021)\citenamefont {Ivanov},
  \citenamefont {Mote}, \citenamefont {Ernst}, \citenamefont {Equbal},\ and\
  \citenamefont {M}}]{Ivanov2021}%
  \BibitemOpen
  \bibfield  {author} {\bibinfo {author} {\bibfnamefont {K.~L.}\ \bibnamefont
  {Ivanov}}, \bibinfo {author} {\bibfnamefont {K.~R.}\ \bibnamefont {Mote}},
  \bibinfo {author} {\bibfnamefont {M.}~\bibnamefont {Ernst}}, \bibinfo
  {author} {\bibfnamefont {A.}~\bibnamefont {Equbal}}, \ and\ \bibinfo {author}
  {\bibfnamefont {P.~K.}\ \bibnamefont {M}},\ }\href {\doibase
  10.1016/j.pnmrs.2021.05.002} {\bibfield  {journal} {\bibinfo  {journal}
  {Prog. Nucl. Magn. Reson. Spectrosc.}\ }\textbf {\bibinfo {volume}
  {126--127}},\ \bibinfo {pages} {17} (\bibinfo {year} {2021})}\BibitemShut
  {NoStop}%
\bibitem [{\citenamefont {Son}\ \emph {et~al.}(2009)\citenamefont {Son},
  \citenamefont {Han},\ and\ \citenamefont {Chu}}]{Son2009}%
  \BibitemOpen
  \bibfield  {author} {\bibinfo {author} {\bibfnamefont {S.~K.}\ \bibnamefont
  {Son}}, \bibinfo {author} {\bibfnamefont {S.}~\bibnamefont {Han}}, \ and\
  \bibinfo {author} {\bibfnamefont {S.~I.}\ \bibnamefont {Chu}},\ }\href
  {\doibase 10.1103/PhysRevA.79.032301} {\bibfield  {journal} {\bibinfo
  {journal} {Phys. Rev. A}\ }\textbf {\bibinfo {volume} {79}},\ \bibinfo
  {pages} {1} (\bibinfo {year} {2009})}\BibitemShut {NoStop}%
\bibitem [{\citenamefont {Hausinger}\ and\ \citenamefont
  {Grifoni}(2010)}]{Hausinger2010}%
  \BibitemOpen
  \bibfield  {author} {\bibinfo {author} {\bibfnamefont {J.}~\bibnamefont
  {Hausinger}}\ and\ \bibinfo {author} {\bibfnamefont {M.}~\bibnamefont
  {Grifoni}},\ }\href {\doibase 10.1103/PhysRevA.81.022117} {\bibfield
  {journal} {\bibinfo  {journal} {Phys. Rev. A}\ }\textbf {\bibinfo {volume}
  {81}},\ \bibinfo {pages} {22117} (\bibinfo {year} {2010})}\BibitemShut
  {NoStop}%
\bibitem [{\citenamefont {Deng}\ \emph {et~al.}(2015)\citenamefont {Deng},
  \citenamefont {Orgiazzi}, \citenamefont {Shen}, \citenamefont {Ashhab},\ and\
  \citenamefont {Lupascu}}]{Deng2015}%
  \BibitemOpen
  \bibfield  {author} {\bibinfo {author} {\bibfnamefont {C.}~\bibnamefont
  {Deng}}, \bibinfo {author} {\bibfnamefont {J.-L.}\ \bibnamefont {Orgiazzi}},
  \bibinfo {author} {\bibfnamefont {F.}~\bibnamefont {Shen}}, \bibinfo {author}
  {\bibfnamefont {S.}~\bibnamefont {Ashhab}}, \ and\ \bibinfo {author}
  {\bibfnamefont {A.}~\bibnamefont {Lupascu}},\ }\href {\doibase
  10.1103/PhysRevLett.115.133601} {\bibfield  {journal} {\bibinfo  {journal}
  {Phys. Rev. Let.}\ }\textbf {\bibinfo {volume} {115}},\ \bibinfo {pages}
  {133601} (\bibinfo {year} {2015})}\BibitemShut {NoStop}%
\bibitem [{\citenamefont {Han}\ \emph {et~al.}(2019)\citenamefont {Han},
  \citenamefont {Luo}, \citenamefont {Li}, \citenamefont {Zhang}, \citenamefont
  {Wang}, \citenamefont {Tsai}, \citenamefont {Nori},\ and\ \citenamefont
  {You}}]{Han2019}%
  \BibitemOpen
  \bibfield  {author} {\bibinfo {author} {\bibfnamefont {Y.}~\bibnamefont
  {Han}}, \bibinfo {author} {\bibfnamefont {X.-Q.}\ \bibnamefont {Luo}},
  \bibinfo {author} {\bibfnamefont {T.-F.}\ \bibnamefont {Li}}, \bibinfo
  {author} {\bibfnamefont {W.}~\bibnamefont {Zhang}}, \bibinfo {author}
  {\bibfnamefont {S.-P.}\ \bibnamefont {Wang}}, \bibinfo {author}
  {\bibfnamefont {J.~S.}\ \bibnamefont {Tsai}}, \bibinfo {author}
  {\bibfnamefont {F.}~\bibnamefont {Nori}}, \ and\ \bibinfo {author}
  {\bibfnamefont {J.~Q.}\ \bibnamefont {You}},\ }\href {\doibase
  10.1103/PhysRevApplied.11.014053} {\bibfield  {journal} {\bibinfo  {journal}
  {Phys. Rev. Appl.}\ }\textbf {\bibinfo {volume} {11}},\ \bibinfo {pages}
  {14053} (\bibinfo {year} {2019})}\BibitemShut {NoStop}%
\bibitem [{\citenamefont {Wang}\ \emph
  {et~al.}(2021{\natexlab{a}})\citenamefont {Wang}, \citenamefont {Liu},\ and\
  \citenamefont {Cappellaro}}]{Wang2021a}%
  \BibitemOpen
  \bibfield  {author} {\bibinfo {author} {\bibfnamefont {G.}~\bibnamefont
  {Wang}}, \bibinfo {author} {\bibfnamefont {Y.-X.}\ \bibnamefont {Liu}}, \
  and\ \bibinfo {author} {\bibfnamefont {P.}~\bibnamefont {Cappellaro}},\
  }\href {\doibase 10.1103/PhysRevA.103.022415} {\bibfield  {journal} {\bibinfo
   {journal} {Phys. Rev. A}\ }\textbf {\bibinfo {volume} {103}},\ \bibinfo
  {pages} {22415} (\bibinfo {year} {2021}{\natexlab{a}})}\BibitemShut {NoStop}%
\bibitem [{\citenamefont {Nishimura}\ \emph {et~al.}(2022)\citenamefont
  {Nishimura}, \citenamefont {Itoh}, \citenamefont {Ishi-Hayase}, \citenamefont
  {Sasaki},\ and\ \citenamefont {Kobayashi}}]{Nishimura2022}%
  \BibitemOpen
  \bibfield  {author} {\bibinfo {author} {\bibfnamefont {S.}~\bibnamefont
  {Nishimura}}, \bibinfo {author} {\bibfnamefont {K.~M.}\ \bibnamefont {Itoh}},
  \bibinfo {author} {\bibfnamefont {J.}~\bibnamefont {Ishi-Hayase}}, \bibinfo
  {author} {\bibfnamefont {K.}~\bibnamefont {Sasaki}}, \ and\ \bibinfo {author}
  {\bibfnamefont {K.}~\bibnamefont {Kobayashi}},\ }\href {\doibase
  10.1103/PhysRevApplied.18.064023} {\bibfield  {journal} {\bibinfo  {journal}
  {Phys. Rev. Appl.}\ }\textbf {\bibinfo {volume} {18}},\ \bibinfo {pages}
  {64023} (\bibinfo {year} {2022})}\BibitemShut {NoStop}%
\bibitem [{\citenamefont {Martin}\ \emph {et~al.}(2017)\citenamefont {Martin},
  \citenamefont {Refael},\ and\ \citenamefont {Halperin}}]{Martin2017}%
  \BibitemOpen
  \bibfield  {author} {\bibinfo {author} {\bibfnamefont {I.}~\bibnamefont
  {Martin}}, \bibinfo {author} {\bibfnamefont {G.}~\bibnamefont {Refael}}, \
  and\ \bibinfo {author} {\bibfnamefont {B.}~\bibnamefont {Halperin}},\ }\href
  {\doibase 10.1103/PhysRevX.7.041008} {\bibfield  {journal} {\bibinfo
  {journal} {Phys. Rev. X}\ }\textbf {\bibinfo {volume} {7}},\ \bibinfo {pages}
  {41008} (\bibinfo {year} {2017})}\BibitemShut {NoStop}%
\bibitem [{\citenamefont {Chen}\ \emph {et~al.}(2020)\citenamefont {Chen},
  \citenamefont {Liu}, \citenamefont {Yu}, \citenamefont {Zhang},\ and\
  \citenamefont {Cai}}]{Chen2020}%
  \BibitemOpen
  \bibfield  {author} {\bibinfo {author} {\bibfnamefont {Q.}~\bibnamefont
  {Chen}}, \bibinfo {author} {\bibfnamefont {H.}~\bibnamefont {Liu}}, \bibinfo
  {author} {\bibfnamefont {M.}~\bibnamefont {Yu}}, \bibinfo {author}
  {\bibfnamefont {S.}~\bibnamefont {Zhang}}, \ and\ \bibinfo {author}
  {\bibfnamefont {J.}~\bibnamefont {Cai}},\ }\href {\doibase
  10.1103/PhysRevA.102.052606} {\bibfield  {journal} {\bibinfo  {journal}
  {Phys. Rev. A}\ }\textbf {\bibinfo {volume} {102}},\ \bibinfo {pages} {52606}
  (\bibinfo {year} {2020})}\BibitemShut {NoStop}%
\bibitem [{\citenamefont {Ikeda}\ and\ \citenamefont {Sato}(2020)}]{Ikeda2020}%
  \BibitemOpen
  \bibfield  {author} {\bibinfo {author} {\bibfnamefont {T.~N.}\ \bibnamefont
  {Ikeda}}\ and\ \bibinfo {author} {\bibfnamefont {M.}~\bibnamefont {Sato}},\
  }\href {\doibase 10.1126/sciadv.abb4019} {\bibfield  {journal} {\bibinfo
  {journal} {Sci. Adv.}\ }\textbf {\bibinfo {volume} {6}},\ \bibinfo {pages}
  {1} (\bibinfo {year} {2020})}\BibitemShut {NoStop}%
\bibitem [{\citenamefont {Ikeda}\ \emph {et~al.}(2021)\citenamefont {Ikeda},
  \citenamefont {Chinzei},\ and\ \citenamefont {Sato}}]{Ikeda2021}%
  \BibitemOpen
  \bibfield  {author} {\bibinfo {author} {\bibfnamefont {T.}~\bibnamefont
  {Ikeda}}, \bibinfo {author} {\bibfnamefont {K.}~\bibnamefont {Chinzei}}, \
  and\ \bibinfo {author} {\bibfnamefont {M.}~\bibnamefont {Sato}},\ }\href
  {\doibase 10.21468/SCIPOSTPHYSCORE.4.4.033/PDF} {\bibfield  {journal}
  {\bibinfo  {journal} {SciPost Phys. Core}\ }\textbf {\bibinfo {volume} {4}},\
  \bibinfo {pages} {033} (\bibinfo {year} {2021})}\BibitemShut {NoStop}%
\bibitem [{\citenamefont {Mori}(2023)}]{Mori2023}%
  \BibitemOpen
  \bibfield  {author} {\bibinfo {author} {\bibfnamefont {T.}~\bibnamefont
  {Mori}},\ }\href {\doibase 10.1146/annurev-conmatphys-040721-015537}
  {\bibfield  {journal} {\bibinfo  {journal} {Annu. Rev. Condens. Matter
  Phys.}\ }\textbf {\bibinfo {volume} {14}},\ \bibinfo {pages} {null} (\bibinfo
  {year} {2023})}\BibitemShut {NoStop}%
\bibitem [{\citenamefont {Shin}\ \emph {et~al.}(2013)\citenamefont {Shin},
  \citenamefont {Avalos}, \citenamefont {Butler}, \citenamefont {Wang},
  \citenamefont {Seltzer}, \citenamefont {Liu}, \citenamefont {Pines},\ and\
  \citenamefont {Bajaj}}]{Shin2013}%
  \BibitemOpen
  \bibfield  {author} {\bibinfo {author} {\bibfnamefont {C.~S.}\ \bibnamefont
  {Shin}}, \bibinfo {author} {\bibfnamefont {C.~E.}\ \bibnamefont {Avalos}},
  \bibinfo {author} {\bibfnamefont {M.~C.}\ \bibnamefont {Butler}}, \bibinfo
  {author} {\bibfnamefont {H.-J.}\ \bibnamefont {Wang}}, \bibinfo {author}
  {\bibfnamefont {S.~J.}\ \bibnamefont {Seltzer}}, \bibinfo {author}
  {\bibfnamefont {R.-B.}\ \bibnamefont {Liu}}, \bibinfo {author} {\bibfnamefont
  {A.}~\bibnamefont {Pines}}, \ and\ \bibinfo {author} {\bibfnamefont {V.~S.}\
  \bibnamefont {Bajaj}},\ }\href {\doibase 10.1103/PhysRevB.88.161412}
  {\bibfield  {journal} {\bibinfo  {journal} {Phys. Rev. B}\ }\textbf {\bibinfo
  {volume} {88}},\ \bibinfo {pages} {161412} (\bibinfo {year}
  {2013})}\BibitemShut {NoStop}%
\bibitem [{\citenamefont {Matsuzaki}\ \emph {et~al.}(2016)\citenamefont
  {Matsuzaki}, \citenamefont {Morishita}, \citenamefont {Shimooka},
  \citenamefont {Tashima}, \citenamefont {Kakuyanagi}, \citenamefont {Semba},
  \citenamefont {Munro}, \citenamefont {Yamaguchi}, \citenamefont {Mizuochi},\
  and\ \citenamefont {Saito}}]{Matsuzaki2016}%
  \BibitemOpen
  \bibfield  {author} {\bibinfo {author} {\bibfnamefont {Y.}~\bibnamefont
  {Matsuzaki}}, \bibinfo {author} {\bibfnamefont {H.}~\bibnamefont
  {Morishita}}, \bibinfo {author} {\bibfnamefont {T.}~\bibnamefont {Shimooka}},
  \bibinfo {author} {\bibfnamefont {T.}~\bibnamefont {Tashima}}, \bibinfo
  {author} {\bibfnamefont {K.}~\bibnamefont {Kakuyanagi}}, \bibinfo {author}
  {\bibfnamefont {K.}~\bibnamefont {Semba}}, \bibinfo {author} {\bibfnamefont
  {W.~J.}\ \bibnamefont {Munro}}, \bibinfo {author} {\bibfnamefont
  {H.}~\bibnamefont {Yamaguchi}}, \bibinfo {author} {\bibfnamefont
  {N.}~\bibnamefont {Mizuochi}}, \ and\ \bibinfo {author} {\bibfnamefont
  {S.}~\bibnamefont {Saito}},\ }\href {\doibase 10.1088/0953-8984/28/27/275302}
  {\bibfield  {journal} {\bibinfo  {journal} {J. Condens. Matter Phys.}\
  }\textbf {\bibinfo {volume} {28}},\ \bibinfo {pages} {275302} (\bibinfo
  {year} {2016})}\BibitemShut {NoStop}%
\bibitem [{\citenamefont {Morishita}\ \emph {et~al.}(2019)\citenamefont
  {Morishita}, \citenamefont {Tashima}, \citenamefont {Mima}, \citenamefont
  {Kato}, \citenamefont {Makino}, \citenamefont {Yamasaki}, \citenamefont
  {Fujiwara},\ and\ \citenamefont {Mizuochi}}]{Morishita2019}%
  \BibitemOpen
  \bibfield  {author} {\bibinfo {author} {\bibfnamefont {H.}~\bibnamefont
  {Morishita}}, \bibinfo {author} {\bibfnamefont {T.}~\bibnamefont {Tashima}},
  \bibinfo {author} {\bibfnamefont {D.}~\bibnamefont {Mima}}, \bibinfo {author}
  {\bibfnamefont {H.}~\bibnamefont {Kato}}, \bibinfo {author} {\bibfnamefont
  {T.}~\bibnamefont {Makino}}, \bibinfo {author} {\bibfnamefont
  {S.}~\bibnamefont {Yamasaki}}, \bibinfo {author} {\bibfnamefont
  {M.}~\bibnamefont {Fujiwara}}, \ and\ \bibinfo {author} {\bibfnamefont
  {N.}~\bibnamefont {Mizuochi}},\ }\href {\doibase 10.1038/s41598-019-49683-z}
  {\bibfield  {journal} {\bibinfo  {journal} {Sci. Rep.}\ }\textbf {\bibinfo
  {volume} {9}},\ \bibinfo {pages} {13318} (\bibinfo {year}
  {2019})}\BibitemShut {NoStop}%
\bibitem [{\citenamefont {Childress}\ and\ \citenamefont
  {McIntyre}(2010)}]{Childress2010}%
  \BibitemOpen
  \bibfield  {author} {\bibinfo {author} {\bibfnamefont {L.}~\bibnamefont
  {Childress}}\ and\ \bibinfo {author} {\bibfnamefont {J.}~\bibnamefont
  {McIntyre}},\ }\href {\doibase 10.1103/PhysRevA.82.033839} {\bibfield
  {journal} {\bibinfo  {journal} {Phys. Rev. A}\ }\textbf {\bibinfo {volume}
  {82}},\ \bibinfo {pages} {33839} (\bibinfo {year} {2010})}\BibitemShut
  {NoStop}%
\bibitem [{\citenamefont {Meinel}\ \emph {et~al.}(2021)\citenamefont {Meinel},
  \citenamefont {Vorobyov}, \citenamefont {Yavkin}, \citenamefont {Dasari},
  \citenamefont {Sumiya}, \citenamefont {Onoda}, \citenamefont {Isoya},\ and\
  \citenamefont {Wrachtrup}}]{Meinel2021}%
  \BibitemOpen
  \bibfield  {author} {\bibinfo {author} {\bibfnamefont {J.}~\bibnamefont
  {Meinel}}, \bibinfo {author} {\bibfnamefont {V.}~\bibnamefont {Vorobyov}},
  \bibinfo {author} {\bibfnamefont {B.}~\bibnamefont {Yavkin}}, \bibinfo
  {author} {\bibfnamefont {D.}~\bibnamefont {Dasari}}, \bibinfo {author}
  {\bibfnamefont {H.}~\bibnamefont {Sumiya}}, \bibinfo {author} {\bibfnamefont
  {S.}~\bibnamefont {Onoda}}, \bibinfo {author} {\bibfnamefont
  {J.}~\bibnamefont {Isoya}}, \ and\ \bibinfo {author} {\bibfnamefont
  {J.}~\bibnamefont {Wrachtrup}},\ }\href {\doibase 10.1038/s41467-021-22714-y}
  {\bibfield  {journal} {\bibinfo  {journal} {Nat. Commun.}\ }\textbf {\bibinfo
  {volume} {12}},\ \bibinfo {pages} {2737} (\bibinfo {year}
  {2021})}\BibitemShut {NoStop}%
\bibitem [{\citenamefont {Han}\ \emph {et~al.}(2020)\citenamefont {Han},
  \citenamefont {Luo}, \citenamefont {Li},\ and\ \citenamefont
  {Zhang}}]{Han2020}%
  \BibitemOpen
  \bibfield  {author} {\bibinfo {author} {\bibfnamefont {Y.}~\bibnamefont
  {Han}}, \bibinfo {author} {\bibfnamefont {X.~Q.}\ \bibnamefont {Luo}},
  \bibinfo {author} {\bibfnamefont {T.~F.}\ \bibnamefont {Li}}, \ and\ \bibinfo
  {author} {\bibfnamefont {W.}~\bibnamefont {Zhang}},\ }\href {\doibase
  10.1103/PhysRevA.101.022108} {\bibfield  {journal} {\bibinfo  {journal}
  {Phys. Rev. A}\ }\textbf {\bibinfo {volume} {101}},\ \bibinfo {pages}
  {022108} (\bibinfo {year} {2020})}\BibitemShut {NoStop}%
\bibitem [{\citenamefont {Leskes}\ \emph {et~al.}(2010)\citenamefont {Leskes},
  \citenamefont {Madhu},\ and\ \citenamefont {Vega}}]{Leskes2010}%
  \BibitemOpen
  \bibfield  {author} {\bibinfo {author} {\bibfnamefont {M.}~\bibnamefont
  {Leskes}}, \bibinfo {author} {\bibfnamefont {P.~K.}\ \bibnamefont {Madhu}}, \
  and\ \bibinfo {author} {\bibfnamefont {S.}~\bibnamefont {Vega}},\ }\href
  {\doibase 10.1016/j.pnmrs.2010.06.002} {\bibfield  {journal} {\bibinfo
  {journal} {Prog. Nucl. Magn. Reson. Spectrosc.}\ }\textbf {\bibinfo {volume}
  {57}},\ \bibinfo {pages} {345} (\bibinfo {year} {2010})}\BibitemShut
  {NoStop}%
\bibitem [{\citenamefont {Wang}\ \emph
  {et~al.}(2021{\natexlab{b}})\citenamefont {Wang}, \citenamefont {Liu},
  \citenamefont {Zhu},\ and\ \citenamefont {Cappellaro}}]{Wang2021b}%
  \BibitemOpen
  \bibfield  {author} {\bibinfo {author} {\bibfnamefont {G.}~\bibnamefont
  {Wang}}, \bibinfo {author} {\bibfnamefont {Y.-X.}\ \bibnamefont {Liu}},
  \bibinfo {author} {\bibfnamefont {Y.}~\bibnamefont {Zhu}}, \ and\ \bibinfo
  {author} {\bibfnamefont {P.}~\bibnamefont {Cappellaro}},\ }\href {\doibase
  10.1021/acs.nanolett.1c01165} {\bibfield  {journal} {\bibinfo  {journal}
  {Nano Lett.}\ }\textbf {\bibinfo {volume} {21}},\ \bibinfo {pages} {5143}
  (\bibinfo {year} {2021}{\natexlab{b}})}\BibitemShut {NoStop}%
\bibitem [{\citenamefont {Michl}\ \emph {et~al.}(2014)\citenamefont {Michl},
  \citenamefont {Teraji}, \citenamefont {Zaiser}, \citenamefont {Jakobi},
  \citenamefont {Waldherr}, \citenamefont {Dolde}, \citenamefont {Neumann},
  \citenamefont {Doherty}, \citenamefont {Manson}, \citenamefont {Isoya},\ and\
  \citenamefont {Wrachtrup}}]{Michl2014}%
  \BibitemOpen
  \bibfield  {author} {\bibinfo {author} {\bibfnamefont {J.}~\bibnamefont
  {Michl}}, \bibinfo {author} {\bibfnamefont {T.}~\bibnamefont {Teraji}},
  \bibinfo {author} {\bibfnamefont {S.}~\bibnamefont {Zaiser}}, \bibinfo
  {author} {\bibfnamefont {I.}~\bibnamefont {Jakobi}}, \bibinfo {author}
  {\bibfnamefont {G.}~\bibnamefont {Waldherr}}, \bibinfo {author}
  {\bibfnamefont {F.}~\bibnamefont {Dolde}}, \bibinfo {author} {\bibfnamefont
  {P.}~\bibnamefont {Neumann}}, \bibinfo {author} {\bibfnamefont {M.~W.}\
  \bibnamefont {Doherty}}, \bibinfo {author} {\bibfnamefont {N.~B.}\
  \bibnamefont {Manson}}, \bibinfo {author} {\bibfnamefont {J.}~\bibnamefont
  {Isoya}}, \ and\ \bibinfo {author} {\bibfnamefont {J.}~\bibnamefont
  {Wrachtrup}},\ }\href {\doibase 10.1063/1.4868128} {\bibfield  {journal}
  {\bibinfo  {journal} {Appl. Phys. Lett.}\ }\textbf {\bibinfo {volume}
  {104}},\ \bibinfo {pages} {102407} (\bibinfo {year} {2014})}\BibitemShut
  {NoStop}%
\bibitem [{\citenamefont {Lesik}\ \emph {et~al.}(2014)\citenamefont {Lesik},
  \citenamefont {Tetienne}, \citenamefont {Tallaire}, \citenamefont {Achard},
  \citenamefont {Mille}, \citenamefont {Gicquel}, \citenamefont {Roch},\ and\
  \citenamefont {Jacques}}]{Lesik2014}%
  \BibitemOpen
  \bibfield  {author} {\bibinfo {author} {\bibfnamefont {M.}~\bibnamefont
  {Lesik}}, \bibinfo {author} {\bibfnamefont {J.-P.}\ \bibnamefont {Tetienne}},
  \bibinfo {author} {\bibfnamefont {A.}~\bibnamefont {Tallaire}}, \bibinfo
  {author} {\bibfnamefont {J.}~\bibnamefont {Achard}}, \bibinfo {author}
  {\bibfnamefont {V.}~\bibnamefont {Mille}}, \bibinfo {author} {\bibfnamefont
  {A.}~\bibnamefont {Gicquel}}, \bibinfo {author} {\bibfnamefont {J.-F.}\
  \bibnamefont {Roch}}, \ and\ \bibinfo {author} {\bibfnamefont
  {V.}~\bibnamefont {Jacques}},\ }\href {\doibase 10.1063/1.4869103} {\bibfield
   {journal} {\bibinfo  {journal} {Appl. Phys. Lett.}\ }\textbf {\bibinfo
  {volume} {104}},\ \bibinfo {pages} {113107} (\bibinfo {year}
  {2014})}\BibitemShut {NoStop}%
\bibitem [{\citenamefont {Fukui}\ \emph {et~al.}(2014)\citenamefont {Fukui},
  \citenamefont {Doi}, \citenamefont {Miyazaki}, \citenamefont {Miyamoto},
  \citenamefont {Kato}, \citenamefont {Matsumoto}, \citenamefont {Makino},
  \citenamefont {Yamasaki}, \citenamefont {Morimoto}, \citenamefont {Tokuda},
  \citenamefont {Hatano}, \citenamefont {Sakagawa}, \citenamefont {Morishita},
  \citenamefont {Tashima}, \citenamefont {Miwa}, \citenamefont {Suzuki},\ and\
  \citenamefont {Mizuochi}}]{Fukui2014}%
  \BibitemOpen
  \bibfield  {author} {\bibinfo {author} {\bibfnamefont {T.}~\bibnamefont
  {Fukui}}, \bibinfo {author} {\bibfnamefont {Y.}~\bibnamefont {Doi}}, \bibinfo
  {author} {\bibfnamefont {T.}~\bibnamefont {Miyazaki}}, \bibinfo {author}
  {\bibfnamefont {Y.}~\bibnamefont {Miyamoto}}, \bibinfo {author}
  {\bibfnamefont {H.}~\bibnamefont {Kato}}, \bibinfo {author} {\bibfnamefont
  {T.}~\bibnamefont {Matsumoto}}, \bibinfo {author} {\bibfnamefont
  {T.}~\bibnamefont {Makino}}, \bibinfo {author} {\bibfnamefont
  {S.}~\bibnamefont {Yamasaki}}, \bibinfo {author} {\bibfnamefont
  {R.}~\bibnamefont {Morimoto}}, \bibinfo {author} {\bibfnamefont
  {N.}~\bibnamefont {Tokuda}}, \bibinfo {author} {\bibfnamefont
  {M.}~\bibnamefont {Hatano}}, \bibinfo {author} {\bibfnamefont
  {Y.}~\bibnamefont {Sakagawa}}, \bibinfo {author} {\bibfnamefont
  {H.}~\bibnamefont {Morishita}}, \bibinfo {author} {\bibfnamefont
  {T.}~\bibnamefont {Tashima}}, \bibinfo {author} {\bibfnamefont
  {S.}~\bibnamefont {Miwa}}, \bibinfo {author} {\bibfnamefont {Y.}~\bibnamefont
  {Suzuki}}, \ and\ \bibinfo {author} {\bibfnamefont {N.}~\bibnamefont
  {Mizuochi}},\ }\href {\doibase 10.7567/APEX.7.055201} {\bibfield  {journal}
  {\bibinfo  {journal} {Appl. Phys. Express}\ }\textbf {\bibinfo {volume}
  {7}},\ \bibinfo {pages} {55201} (\bibinfo {year} {2014})}\BibitemShut
  {NoStop}%
\bibitem [{\citenamefont {Ishiwata}\ \emph {et~al.}(2017)\citenamefont
  {Ishiwata}, \citenamefont {Nakajima}, \citenamefont {Tahara}, \citenamefont
  {Ozawa}, \citenamefont {Iwasaki},\ and\ \citenamefont
  {Hatano}}]{Ishiwata2017}%
  \BibitemOpen
  \bibfield  {author} {\bibinfo {author} {\bibfnamefont {H.}~\bibnamefont
  {Ishiwata}}, \bibinfo {author} {\bibfnamefont {M.}~\bibnamefont {Nakajima}},
  \bibinfo {author} {\bibfnamefont {K.}~\bibnamefont {Tahara}}, \bibinfo
  {author} {\bibfnamefont {H.}~\bibnamefont {Ozawa}}, \bibinfo {author}
  {\bibfnamefont {T.}~\bibnamefont {Iwasaki}}, \ and\ \bibinfo {author}
  {\bibfnamefont {M.}~\bibnamefont {Hatano}},\ }\href {\doibase
  10.1063/1.4993160} {\bibfield  {journal} {\bibinfo  {journal} {Appl. Phys.
  Lett.}\ }\textbf {\bibinfo {volume} {111}},\ \bibinfo {pages} {43103}
  (\bibinfo {year} {2017})}\BibitemShut {NoStop}%
\bibitem [{\citenamefont {Sasaki}\ \emph {et~al.}(2016)\citenamefont {Sasaki},
  \citenamefont {Monnai}, \citenamefont {Saijo}, \citenamefont {Fujita},
  \citenamefont {Watanabe}, \citenamefont {Ishi-Hayase}, \citenamefont {Itoh},\
  and\ \citenamefont {Abe}}]{Sasaki2016}%
  \BibitemOpen
  \bibfield  {author} {\bibinfo {author} {\bibfnamefont {K.}~\bibnamefont
  {Sasaki}}, \bibinfo {author} {\bibfnamefont {Y.}~\bibnamefont {Monnai}},
  \bibinfo {author} {\bibfnamefont {S.}~\bibnamefont {Saijo}}, \bibinfo
  {author} {\bibfnamefont {R.}~\bibnamefont {Fujita}}, \bibinfo {author}
  {\bibfnamefont {H.}~\bibnamefont {Watanabe}}, \bibinfo {author}
  {\bibfnamefont {J.}~\bibnamefont {Ishi-Hayase}}, \bibinfo {author}
  {\bibfnamefont {K.~M.}\ \bibnamefont {Itoh}}, \ and\ \bibinfo {author}
  {\bibfnamefont {E.}~\bibnamefont {Abe}},\ }\href {\doibase 10.1063/1.4952418}
  {\bibfield  {journal} {\bibinfo  {journal} {Rev. Sci. Instrum.}\ }\textbf
  {\bibinfo {volume} {87}},\ \bibinfo {pages} {053904} (\bibinfo {year}
  {2016})}\BibitemShut {NoStop}%
\bibitem [{\citenamefont {Bloch}\ and\ \citenamefont
  {Siegert}(1940)}]{Bloch1940}%
  \BibitemOpen
  \bibfield  {author} {\bibinfo {author} {\bibfnamefont {F.}~\bibnamefont
  {Bloch}}\ and\ \bibinfo {author} {\bibfnamefont {A.}~\bibnamefont
  {Siegert}},\ }\href {\doibase 10.1103/PhysRev.57.522} {\bibfield  {journal}
  {\bibinfo  {journal} {Phys. Rev.}\ }\textbf {\bibinfo {volume} {57}},\
  \bibinfo {pages} {522} (\bibinfo {year} {1940})}\BibitemShut {NoStop}%
\bibitem [{\citenamefont {Uchida}\ \emph {et~al.}(2022)\citenamefont {Uchida},
  \citenamefont {Kusaba}, \citenamefont {Nagai}, \citenamefont {Ikeda},\ and\
  \citenamefont {Tanaka}}]{Uchida2022}%
  \BibitemOpen
  \bibfield  {author} {\bibinfo {author} {\bibfnamefont {K.}~\bibnamefont
  {Uchida}}, \bibinfo {author} {\bibfnamefont {S.}~\bibnamefont {Kusaba}},
  \bibinfo {author} {\bibfnamefont {K.}~\bibnamefont {Nagai}}, \bibinfo
  {author} {\bibfnamefont {T.~N.}\ \bibnamefont {Ikeda}}, \ and\ \bibinfo
  {author} {\bibfnamefont {K.}~\bibnamefont {Tanaka}},\ }\href {\doibase
  10.1126/sciadv.abq7281} {\bibfield  {journal} {\bibinfo  {journal} {Science
  Advances}\ }\textbf {\bibinfo {volume} {8}},\ \bibinfo {pages} {eabq728}
  (\bibinfo {year} {2022})}\BibitemShut {NoStop}%
\bibitem [{\citenamefont {Tashima}\ \emph {et~al.}(2019)\citenamefont
  {Tashima}, \citenamefont {Morishita},\ and\ \citenamefont
  {Mizuochi}}]{Tashima2019}%
  \BibitemOpen
  \bibfield  {author} {\bibinfo {author} {\bibfnamefont {T.}~\bibnamefont
  {Tashima}}, \bibinfo {author} {\bibfnamefont {H.}~\bibnamefont {Morishita}},
  \ and\ \bibinfo {author} {\bibfnamefont {N.}~\bibnamefont {Mizuochi}},\
  }\href {\doibase 10.1103/PhysRevA.100.023801} {\bibfield  {journal} {\bibinfo
   {journal} {Phys. Rev. A}\ }\textbf {\bibinfo {volume} {100}},\ \bibinfo
  {pages} {23801} (\bibinfo {year} {2019})}\BibitemShut {NoStop}%
\end{thebibliography}
\end{document}